\documentclass[12pt,tightenlines,eqsecnum,floats,aps,amsmath,amssymb,nofootinbib,prd,showpacs]{revtex4}

\usepackage{setspace}
\usepackage{amsmath,amssymb,amsfonts,amsthm}
\usepackage{graphicx}
\usepackage{enumerate} 
\usepackage{colordvi} 

\def\be{\begin{equation}}
\def\ee{\end{equation}}
\def\ba{\begin{eqnarray}}
\def\ea{\end{eqnarray}}

\def\p{\partial}
\def\V{\mathcal{C}}
\def\H{{\cal H}}
\def\Hkg{\H_{\rm kin}^{\rm grav}}
\def\Hk{\H_{\rm kin}}
\def\Hp{\H_{\rm phy}}

\def\h{\hat }
\def\b{\bar}

\def\cm{\rm cm}

\def\pphi{p_{(\phi)}}

\def\lp{{\ell}_{\rm Pl}}

\def\q{{}^o\!q}
\def\e{{}^o\!e}

\def\rcr{\rho_{\mathrm{crit}}}
\def\rmin{\rho_{\mathrm{min}}}
\def\rmax{\rho_{\mathrm{max}}}
\def\b{$\bullet\,\,\,\, $}
\def\f{\frac}

\def\ul{\underline}
\def\WDW{WDW\,\,}

\newcommand{\fs}[2]{{\textstyle\frac{#1}{#2}}} 

\usepackage{enumerate}

\usepackage{colordvi}
\newcounter{mnotecount}[section]

\newcommand{\comment}[1]{}

\begin{document}

\title{Loop Quantum Cosmology: An Overview}

\author{Abhay Ashtekar}
\email{ashtekar@gravity.psu.edu} \affiliation{Institute for
Gravitation and the Cosmos, Physics Department, Penn State,
University Park, PA 16802, U.S.A.}

\begin{abstract}

A brief overview
\footnote{To appear in the Proceedings of the Bad Honef Workshop
entitled \emph{Quantum Gravity: Challenges and Perspectives},
dedicated to the memory of John A. Wheeler.}
of loop quantum cosmology of homogeneous isotropic models is
presented with emphasis on the origin of and subtleties associated
with the resolution of big bang and big crunch singularities. These
results bear out the remarkable intuition that John Wheeler had.
Discussion is organized at two levels. The the main text provides a
bird's eye view of the subject that should be accessible to
non-experts. Appendices address conceptual and technical issues that
are often raised by experts in loop quantum gravity and string
theory.

\end{abstract}


\maketitle

\section{Introduction}
\label{s1}

In general relativity, the gravitational field is encoded in the
very geometry of space-time. Geometry is no longer an inert backdrop
providing just a stage for physical happenings; it is a physical
entity that interacts with matter. As we all know, this deep
paradigm shift lies at the heart of the most profound predictions of
the theory: the big bang, the black holes and the gravitational
waves. However, the encoding of gravity in geometry also implies
that space-time itself must end when the gravitational field becomes
singular, and \emph{all of physics} must come to an abrupt halt.
This is why in popular articles relativists and cosmologists like to
say that the universe was born with a big bang some 14 billion years
ago and it is meaningless to ask what was there before. In more
technical articles they point out that this finite beginning leads
to the `horizon problem.' But we know that general relativity is
incomplete because it ignores quantum physics. If we go back in time
using general relativity, we encounter huge matter densities and
curvatures at which quantum effects should in fact dominate physics.
This is the regime where we can no longer trust general relativity.
\emph{Thus, the big bang is a prediction of general relativity
precisely in a domain where it is inapplicable.} Although in the
framework of general relativity the universe did begin with a
big-bang, there is no reason to believe that the real, physical
universe did.

To know what really happened, one needs a quantum theory of gravity.
John Wheeler recognized this early and drew on various analogies to
argue that quantum fluctuations of geometry would intervene and
resolve classical singularities. Already in his 1967 lectures in
Battelle Rencontres, he wrote \cite{jw1}:
\begin{quote}
{\sl In all applications of quantum geometrodynamics, none would
seem more immediate than gravitational collapse Here, according to
classical general relativity, the dimensions of collapsing system
are driven down to indefinitely small values. 
... In a finite proper time the calculated curvature rises to
infinity. At this point the classical theory becomes incapable of
further prediction. In actuality, classical predictions go wrong
before this point. A prediction of infinity is not a prediction.
The wave packet in superspace does not and cannot follow the
classical history when the geometry becomes smaller in scale than
the quantum mechanical spread of the wave packet. ... The
semiclassical treatment of propagation is appropriate in most of
the domain of superspace of interest to gravitational collapse.
Not so in the decisive region.}
\end{quote}

Wheeler focused on the gravitational collapse but his comments are
equally applicable to the big-bang ---the time reverse of the
collapse. This point was made explicit is Misner's articles on
quantum cosmology \cite{cwm}, particularly in his contribution to
the Wheeler Festschrift.

However, It turned out that within the framework of quantum
geometrodynamics (QGD) that Wheeler, Misner and DeWitt used, without
additional inputs the big bang singularity could not be resolved
generically, at least in the precise, physical sense spelled out in
section \ref{s4}. The subject had therefore remained rather dormant
for over two decades. Over the last 6-7 years, the issue was revived
in the context of loop quantum cosmology (LQC)\cite{mb-rev,aa-rev}
---the application of loop quantum gravity (LQG)
\cite{alrev,crbook,ttbook} to cosmological models. The LQG program
is rather similar to that envisaged by Wheeler: both are canonical
approaches, both follow pioneering ideas of Bergmann and Dirac,
and in both cases dynamics has to be teased out of the quantum
Hamiltonian constraint. There is however, a key difference: LQG is
based on a specific quantum theory of Riemannian geometry. As a
result, geometric observables display a fundamental discreteness
\cite{almmt,rs,al5,alvol}. It turns out that this discreteness
plays a key role in quantum dynamics: While predictions of LQC are
very close to those of QGD away from the Planck regime, there is a
dramatic difference once densities and curvatures enter the Planck
scale. In LQC the big bang is replaced by a quantum bounce.
Moreover, thanks to the introduction of new analytical and
numerical methods over the past two years, it is now possible to
probe the Planck scale physics in detail.

The purpose of this article is to present an overview of the
situation. Since the intended audience is diverse, I will present
the material at two levels. In the main body, I will summarize
the main developments, emphasizing the conceptual aspects%
\footnote{Thus I will not include any derivations but instead
provide references where the details can be found.}
from an angle that, I hope, will make the material accessible also
to non-experts. In the Appendices, I will address a number of more
technical issues that are often raised in loop quantum gravity as
well as string theory circles.

The material is organized as follows. In section \ref{s2} I will set
the stage by listing some key questions on the nature of the big
bang that any quantum theory of gravity should address. In section
\ref{s3}, I will summarize the situation in the Wheeler DeWitt
theory. Because LQC has evolved considerably since its inception,
the cumulative discussion in the literature can seem somewhat
confusing to non-experts. Therefore, I will begin in section
\ref{s4} by outlining this evolution of the subject and then explain
the key difference between LQC and QGD (also known as the
Wheeler-DeWitt (WDW) theory). Main results of LQC are gathered in
section \ref{s5}. Section \ref{s6} summarizes the overall situation
from a broad perspective. Appendix \ref{a1} discusses some
conceptual issues and \ref{a2}, issues related to dynamics.

\section{Some key questions}
\label{s2}

Many of the key questions that any approach to quantum gravity
should address in the cosmological context were already raised in
the seventies by DeWitt, Misner and Wheeler. More recent
developments in inflationary and cyclic models raise additional
issues. In this section, I will present a prototype list. It is far
from being complete but should suffice to provide an approach
independent gauge to compare the status of various programs.

\begin{itemize}
\item How close to the big-bang does a smooth space-time of
    general relativity make sense?  Inflationary scenarios, for
    example, are based on a space-time continuum. Can one show
    from `first principles' that this is a safe approximation
    already at the onset of inflation?

\item Is the big-bang singularity naturally resolved by quantum gravity?
      As we saw in section \ref{s1}, this tantalizing possibility led
      to the development of the field of quantum cosmology in the late
      1960s. The basic idea can be illustrated using an analogy to the
      theory of the hydrogen atom. In classical electrodynamics the ground
      state energy of this system is unbounded below. Quantum physics
      intervenes and, thanks to a non-zero Planck's constant, the
      ground state energy is lifted to a finite value, $-me^4/2\hbar^2
      \approx - 13.6{\rm eV}$. Since it is the Heisenberg uncertainly
      principle that lies at the heart of this resolution and since the
      principle is fundamental also to quantum gravity, one is led to ask:
      Can a similar mechanism resolve the big-bang and big crunch singularities
      of general relativity?

\item Is a new principle/ boundary condition at the big bang or
    the big crunch essential? The most well known example of
    such a boundary condition is the `no boundary proposal' of
    Hartle and Hawking \cite{hh}. Or, do quantum Einstein
    equations suffice by themselves even at the classical
    singularities?

\item Do quantum dynamical equations remain well-behaved even at
    these singularities? If so, do they continue to provide a
    deterministic evolution? The idea that there was a
    pre-big-bang branch to our universe has been advocated in
    several approaches, most notably by the pre-big-bang
    scenario in string theory \cite{pbb} and ekpyrotic and
    cyclic models \cite{ekp1,ekp2} inspired by the brane world
    ideas. However, these are perturbative treatments which
    require a smooth continuum in the background. Therefore,
    their dynamical equations break down at the singularity
    whence, without additional input, the pre-big-bang branch is
    not joined to the current post-big-bang branch by a
    deterministic evolution. Can one improve on this situation?

\item  If there is a deterministic evolution, what is on the
    `other side'? Is there just a quantum foam from which the
    current post-big-bang branch is born, say a `Planck time
    after the putative big-bang'? Or, was there another
    classical universe as in the pre-big-bang and cyclic
    scenarios, joined to ours by deterministic equations?
\end{itemize}

Clearly, to answer such questions we cannot start by assuming that
there is a smooth space-time in the background. But already in the
classical theory, it took physicists several decades to truly
appreciate the dynamical nature of geometry and to learn to do
physics without recourse to a background space-time. In quantum
gravity, this
issue becomes even more vexing.%
\footnote{There is a significant body of literature on issue; see
e.g., \cite{as-book} and references therein. These difficulties are
now being discussed also in the string theory literature in the
context of the AdS/CFT conjecture.}

For simple systems, including Minkowskian field theories, the
Hamiltonian formulation generally serves as the royal road to
quantum theory. It was therefore adopted for quantum gravity by
Dirac, Bergmann, Wheeler and others. But absence of a background
metric implies that the Hamiltonian dynamics is generated by
constraints \cite{kk}. In the quantum theory, physical states are
solutions to quantum constraints. All of physics, including the
dynamical content of the theory, has to be extracted from these
solutions. But there is no external time to phrase questions about
evolution. Therefore we are led to ask:

\begin{itemize}

\item Can we extract, from the arguments of the wave function,
    one variable which can serve as \emph{emergent time} with
    respect to which the other arguments `evolve'? Such an
    internal or emergent time is not essential to obtain a
    complete, self-contained theory. But its availability makes
    the physical meaning of dynamics transparent and one can
    extract the phenomenological predictions more easily. In a
    pioneering work, DeWitt proposed that the determinant of the
    3-metric can be used as internal time \cite{bd}.
    Consequently, in much of the literature on the
    Wheeler-DeWitt (\WDW) approach to quantum cosmology, the
    scale factor is assumed to play the role of time, although
    sometimes only implicitly. However, in closed models the
    scale factor fails to be monotonic due to classical
    recollapse and cannot serve as a global time variable
    already in the classical theory. Are there better
    alternatives at least in the simple setting of quantum
    cosmology?

\end{itemize}

Finally there is an important ultraviolet-infrared tension,
emphasized by Green and Unruh \cite{gu} in the context of an older
LQC treatment of the k=1 model:

\begin{itemize}

\item Can one construct a framework that cures the
    short-distance difficulties faced by classical general
    reltivity near singularities, while maintaining an agreement
    with it at large scales?

\end{itemize}

By their very construction, perturbative and effective
descriptions have no problem with the second requirement. However,
physically their implications can not be trusted at the Planck
scale and mathematically they generally fail to provide a
deterministic evolution across the putative singularity. Since the
non-perturbative approaches often start from deeper ideas, it is
conceivable that they could lead to new structures at the Planck
scale which modify the classical dynamics and resolve the big-bang
singularity. But once unleashed, do these new quantum effects
naturally `turn-off' sufficiently fast, away from the Planck
regime? The universe has had some \emph{14 billion years} to
evolve since the putative big bang and even minutest quantum
corrections could accumulate over this huge time period leading to
observable departures from dynamics predicted by general
relativity. Thus, the challenge to quantum gravity theories is to
first create huge quantum effects that are capable of overwhelming
the extreme gravitational attraction produced by matter densities
of some $10^{94}\, {\rm gms/cc}$ near the big bang, and then
switching them off with extreme rapidity as the matter density
falls below this Planck scale. This is a huge burden!

The question then is: How do various approaches fare with respect to
these questions? In LQC these issues have been addressed in
considerable detail. As we will see in sections \ref{s4} and
\ref{s5}, some of them could be addressed even in the simplest,
early versions of the theory, while others required a much more
careful analysis of the quantum Hamiltonian constraint.

\section{FRW models and the \WDW theory}
 \label{s3}

\begin{figure}[]
  \begin{center}
$a)$\hspace{8cm}$b)$
    \includegraphics[width=3.2in,angle=0]{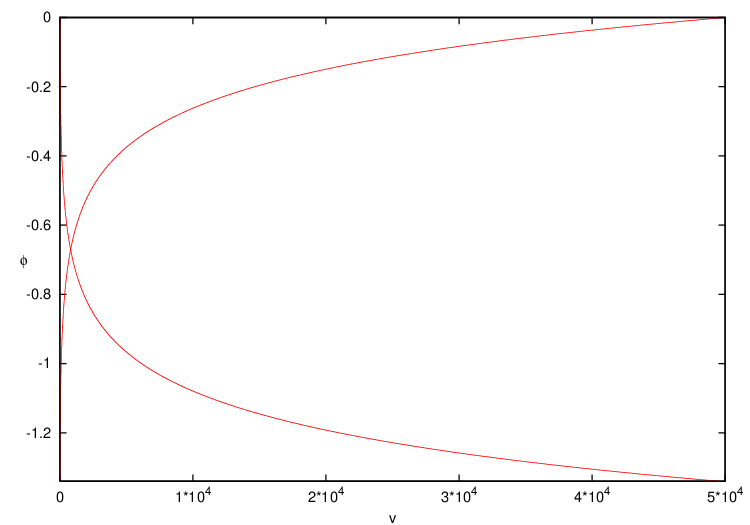}
\includegraphics[width=3.2in,angle=0]{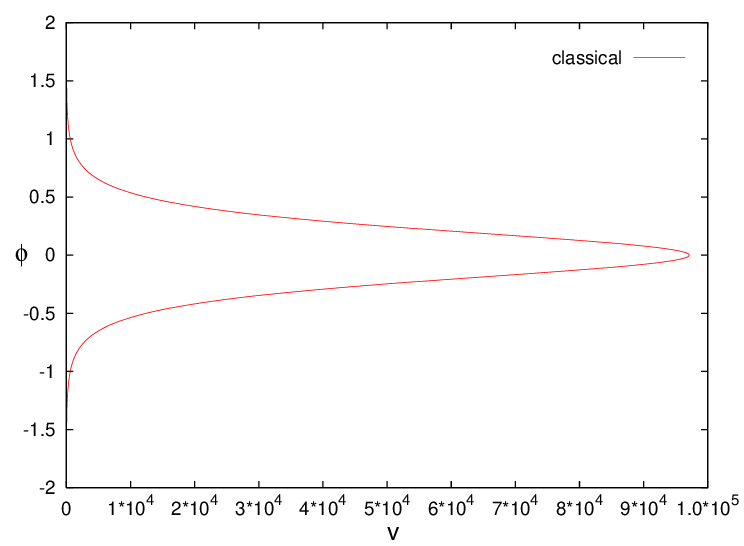}
\caption{$a)$ Classical solutions in k=0, $\Lambda=0$ FRW models
with a massless scalar field. Since $\pphi$ is a constant of motion,
a classical trajectory can be plotted in the $v$-$\phi$ plane, where
$v$ is the volume (essentially in Planck units) of a fixed fiducial
cell. There are two classes of trajectories. In one the universe
begins with a big-bang and expands and in the other it contracts
into a big crunch.\\ $b)$ Classical solutions in the k=1, $\Lambda =0$
FRW model with a massless scalar field. The universe begins with a
big bang, expands to a maximum volume and then undergoes a
recollapse to a big crunch singularity. Since the volume of the
universe is double valued in any solution, it cannot serve as a
global time coordinate in this case. The scalar field on the other
hand does so both in the k=0 and k=1 cases.}\label{class}
  \end{center}
\end{figure}
Almost all phenomenological work in cosmology is based on the k=0
homogeneous and isotropic Friedmann Robertson Walker (FRW)
space-times and perturbations thereof. Therefore, these models
provide a natural point of departure for quantum cosmology. For
concreteness, I will focus on FRW model in which the only matter
source is a massless scalar field (although our discussion will make
it clear that it is relatively straightforward to allow additional
fields, possibly with complicated potentials). I will consider k=0
(or spatially flat) as well as k=1 (spatially closed) models.
Conceptually, these models are interesting for our purpose because
\emph{every} of their classical solutions has a singularity (see Fig
\ref{class}). Therefore a natural singularity resolution without
external inputs is highly non-trivial. In light of the spectacular
observational inputs over the past decade, the k=0 model is the one
that is phenomenologically most relevant. However as we will see,
because of its classical recollapse, the k=1 model offers a more
stringent viability test for the quantum cosmology.

Let us begin with the issue of time. In the classical theory, one
considers one space-time at a time and although the metric of that
space-time is dynamical, it enables one to introduce time
coordinates ---such as the proper time-- that have direct physical
significance. However in the quantum theory
---and indeed already in the phase space framework that serves as
the stepping stone to quantum theory--- we have to consider all
possible homogeneous, isotropic space-times. In this setting one can
introduce a natural foliation of the 4-manifold, each leaf of which
serves as the `home' to a spatially homogeneous 3-geometry. However,
unlike in non-gravitational theories, there is no preferred physical
\emph{time variable} to define evolution. As discussed in section
\ref{s2}, a natural strategy is to use part of the system as an
`internal' clock with respect to which the rest of the system
evolves. This leads one to Leibnitz's \emph{relational time}. Now,
in any spatially homogeneous model with a massless scalar field
$\phi$, the conjugate momentum $p_{(\phi)}$ is a constant of motion,
whence $\phi$ is monotonic along any dynamical trajectory. Thus, in
the classical theory, it serves as a
global clock%
\footnote{Although $\phi$ is a good evolution parameter, it does not
have physical dimensions of time. Still in what follows we will
loosely refer to it as `time' for simplicity.}
(see Fig. \ref{class}). Questions about evolution can thus be
phrased as: ``If the curvature or matter density or an anisotropy
parameter is such and such when $\phi =\phi_1$ what is it when
$\phi = \phi_2$?''

What is the situation in the quantum theory? There is no a priori
guarantee that a variable which serves as a viable time parameter in
the classical theory will continue to do so in the quantum theory.
Whether it does so depends on the form of the Hamiltonian
constraint. For instance as Fig \ref{class}a shows, in the k=0 model
without a cosmological constant, volume (or the scale factor) can be
used as a global `clock' along any classical trajectory. But the
form of the quantum Hamiltonian constraint \cite{aps2} in loop
quantum gravity is such that it does not readily serve this role in
the quantum theory. The scalar field $\phi$, on the other hand,
continues to do so (with or without a cosmological
constant and also in the k=1 case).%
\footnote{As I mentioned in section \ref{s2}, while the availability
of an internal evolution parameter such as $\phi$ makes it easier to
interpret the theory, a preferred time variable,
---especially one that is defined globally--- is not essential.
If there is no massless scalar field, one could still use a
suitable matter field as a `local' internal clock. For instance,
in the inflationary scenario, because of the presence of the
potential, the inflaton is not monotonic even along classical
trajectories. But it is possible to divide dynamics into `epochs'
and use the inflaton as a clock locally, i.e., within each epoch
\cite{aps4}. There is considerable literature on the issue of
internal time for model constrained systems \cite{as-book} (such
as a system of two harmonic oscillators where the total energy is
constrained to be constant \cite{cr-time}).}

Let us now turn to quantization. Because of the assumption of
spatial homogeneity, we have only a finite number of degrees of
freedom. Therefore, although the conceptual problems of quantum
gravity remain, there are no field theoretical infinities and one
can hope to mimic ordinary text book quantum mechanics to pass to
quantum theory. However, in the k=0 case, because space is infinite,
homogeneity implies that the action, the symplectic structure and
Hamiltonians all diverge since they are represented as integrals
over all of space. Therefore, in any approach to quantum cosmology
---irrespective of whether it is based on path integrals or
canonical methods--- one has to introduce an elementary cell $\V$
and restrict all integrals to it. In actual calculations, it is
generally convenient also to introduce a fiducial 3-metric $\q_{ab}$
(as well as frames $\e^a_i$ adapted to the spatial isometries) and
represent the physical metric $q_{ab}$ via a scale factor $a$, \,
$q_{ab} = a^2\, \q_{ab}$. Then the geometrical dynamical variable
can be taken to be either the scale factor $a$ or the `oriented'
volume $v$ of the fiducial cell $\V$ as measured by the physical
frame $e^a_i$, where $v$ is positive if $e^a_i$ has the same
orientation as $\e^a_i$ and negative if the orientations are
opposite. (In either case the physical volume of the cell is $|v|$.)
In LQC it is more convenient to use $v$ rather than the scale factor
so we will use $v$ here as well. Note, however, that physical
results cannot depend on the choice of the fiducial cell $\V$ or
the fiducial metric $\q_{ab}$.%
\footnote{This may appear as an obvious requirement but
unfortunately it is often overlooked in the literature. The claimed
physical results often depend on the choice of $\V$ and/or $\q_{ab}$
although the dependence is often hidden by setting the volume $v_o$
of $\V$ with respect to $\q_{ab}$ to $1$ (in unspecified units) in
the classical theory.}
In the k=1 case, since space is compact, a fiducial cell is
unnecessary and the dynamical variable $v$ is then just the oriented
physical volume of the universe.

With this caveat about the elementary cell out of the way, one can
proceed with quantization. Situation in the \WDW theory can be
summarized as follows. This theory emerged in the late sixties and
was analyzed extensively over the next decade and a half \cite{kk}.
Many of the key physical ideas of quantum cosmology were introduced
during this period \cite{bd,cwm} and a number of models were
analyzed. However, since a mathematically coherent approach to
quantization of full general relativity did not exist, there were no
guiding principles for the analysis of these simpler, symmetry
reduced systems. Rather, quantization was carried out following
`obvious' methods from ordinary quantum mechanics. Thus, in quantum
kinematics, states were represented by square integrable wave
functions $\Psi(v,\phi)$, where $v$ represents geometry and $\phi$,
matter; and operators $\hat{v}, \hat\phi$ acted by multiplication
and their conjugate momenta by ($-i \hbar$ times) differentiation.

Because of spatial homogeneity and isotropy, we are left with a
single Hamiltonian constraint; all others are automatically
solved. The Hamiltonian constraint takes the form of a
differential equation that must be satisfied by the physical
states \cite{aps3}:
\be \label{wdw0} \p_\phi^2 \ul\Psi(v,\phi) =
{\ul{\Theta}}_o\ul\Psi(v,\phi)\, := \, -12\pi G\, (v\partial_v)^2\,
\ul\Psi(v,\phi)\, \ee
for k=0, and
\be \label{wdw1} \p_\phi^2 \ul\Psi(v,\phi) = -{\ul{\Theta}}_1
\ul\Psi(v,\phi) \, := \, -{\ul{\Theta}}_o \ul\Psi(v,\phi) -
G\,C\,|v|^{\frac{4}{3}}\, \ul\Psi(v,\phi) \ , \ee
for k=1, where $C$ is a numerical constant. (The bars below various
symbols indicate that they refer to the \WDW theory.) \emph{In what
follows $\ul{\Theta}$ will stand for either $\ul{\Theta}_o$ or
$\ul{\Theta}_1$.} These are the celebrated \WDW equations of the two
models. In the older literature, the emphasis was on finding and
interpreting the WKB solutions of these equations (see, e.g.,
\cite{ck}). However, as Wheeler emphasized already in 1968
\cite{jw1}, the WKB approximation fails near the singularity and we
need an exact quantum theory.

The physical sector of the \WDW theory can be readily constructed
\cite{aps2,aps3}. A systematic procedure based on the so-called
group averaging method \cite{dm} (which is applicable for a very
large class of constrained systems) provides the physical inner
product on the space of solutions $\Psi(v,\phi)$ to the \WDW
equation. To understand its structure, note that the form of
(\ref{wdw0}) and (\ref{wdw1}) is the same as that of a
Klein-Gordon equation in a 2-dimensional static space-time (with a
$\phi$-independent potential in the k=1 case), where $\phi$ plays
the role of time and $v$ of space. This suggests that we think of
$\phi$ as the relational time variable with respect to which $v$,
the `true' degree of freedom, evolves. Not surprisingly, the
scalar product given by the group averaging procedure coincides
with the expression from the Klein-Gordon theory in static
space-times.

\begin{figure}[]
  \begin{center}
    \includegraphics[width=3.2in,angle=0]{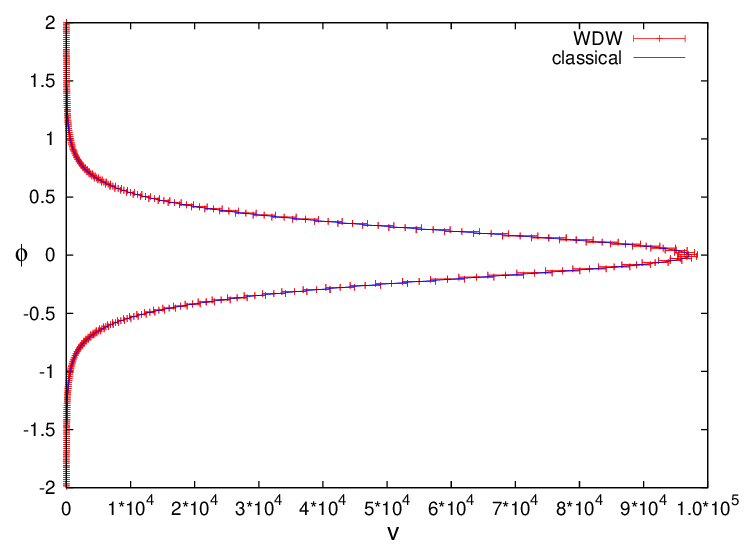}
\caption{ Expectation values (and dispersions) of
${|\hat{v}|_{\phi}}$ for the \WDW wave function in the k=1 model.
The \WDW wave function follows the classical trajectory into the
big-bang and big-crunch singularities. ($p_{(\phi)}$ is a constant
of motion. In this simulation, the quantum state at late times is
a Gaussian peaked at $p_{(\phi)} = 5000 \hbar$ in the G=c=1 units and
the dispersion is $\Delta p_{(\phi)}/p_{(\phi)} = 0.02$.) }
    \label{fig:wdw}
 \end{center}
\end{figure}

The resulting physical sector of the final theory can be summarized
as follows. The physical Hilbert space $\Hp$ in the k=0 and k=1
cases consists of `positive frequency' solutions to (\ref{wdw0}) and
(\ref{wdw1}) respectively. A complete set of observables is provided
by the momentum $\hat{p}_{(\phi)}$ and the relational observable
$\hat{|v|}|_{\phi_o}$ representing the volume at the `instant of
time $\phi_o$':
\be \h{p}_{(\phi)}\, \underline{\Psi}(v,\phi) = -i\hbar
\partial_\phi\, \underline{\Psi}(v,\phi)\quad{\rm and}\quad
\h{V}|_{\phi_o}\, \underline{\Psi}(v,\phi) = e^{i
\,\sqrt{\ul{\Theta}}\,(\phi-\phi_o)}\, |v|\, \Psi(v,\phi_o)\,. \ee
(Thus, the action of $\hat{|v|}|_{\phi_o}$ is as follows: One
freezes the given solution to the \WDW equation at `time' $\phi_o$,
acts of the volume operator and then evolves the new initial data to
obtain another solution.) There are \emph{Dirac} observables because
their action preserves the space of solutions to the constraints and
are self-adjoint on the physical Hilbert space $\Hp$. With this
exact quantum theory at hand, we can ask if the singularities are
naturally resolved.

More precisely, from $\h{p}_{(\phi)}$ and $\h{V}|_\phi$ we can
construct observables corresponding to matter density $\h\rho$ (or
space-time scalar curvature $\h{R}$). Since the singularity is
characterized by divergence of these quantities in the classical
theory, in the quantum theory we can proceed as follows. We can
select a point $(v_o, \phi_o)$ at a `late time' $\phi_o$ on a
classical trajectory of Fig \ref{class} ---e.g., the present epoch
in the history of our universe--- when the density and curvature are
\emph{very} low compared to the Planck scale, and construct a
semi-classical state which is sharply peaked at $v_o$ at $\phi
=\phi_o$. We can then evolve this state \emph{backward} in time.
Does it follow the classical trajectory? To have the correct
`infra-red' behavior, it must, until the density and curvature
become very high. What happens in this `ultra-violet' regime? Does
the quantum state remain semi-classical and follow the classical
trajectory into the big bang? Or, does it spread out making quantum
fluctuations so large that although the quantum evolution does not
break down, there is no reasonable notion of classical geometry? Or,
does it remain peaked on some trajectory which however is so
different from the classical one that, in this backward evolution,
the universe `bounces' rather than being crushed into the
singularity? Or, does it ... Each of these scenarios provides a
distinct prediction for the ultra-violet behavior and therefore for
physics in the deep Planck regime.%
\footnote{Sometimes apparently weaker notions of singularity
resolution are discussed. Consider two examples \cite{kks}. One may
be able to show that the wave function vanishes at points of the
classically singular regions of the configuration space. However,
the \emph{physical} inner product could well be  non-local in this
configuration space, whence such a behavior of the wave function
would not imply that the probability of finding the universe at
these configurations is zero. The second example is that the wave
function may become highly non-classical. This by itself would not
mean that the singularity is avoided unless one can show that the
expectation values of a family of Dirac observables which become
classically singular remain finite there.}

It turns out that the \WDW theory leads to similar predictions in
both k=0 and k=1 cases \cite{aps2,aps3,apsv,warsaw1}. They pass the
infra-red tests with flying colors (see Fig \ref{fig:wdw}). But
unfortunately the state follows the classical trajectory into the
big bang (and in the k=1 case also the big crunch) singularity. Thus
the first of the possibilities listed above is realized. The
singularity is not resolved because expectation values of density
and curvature continue to diverge in epochs when their classical
counterparts do. The analogy to the hydrogen atom discussed in
section \ref{s2} fails to be realized in the \WDW theory of these
simple models.

\section{LQC: Preliminaries}
 \label{s4}

For a number of years, the failure of the \WDW theory to naturally
resolve the big bang singularity was taken to mean that quantum
cosmology cannot, by itself, shed significant light on the quantum
nature of the big bang. Indeed, for systems with a finite number
of degrees of freedom we have the von Neumann uniqueness theorem
which guarantees that quantum kinematics is unique. The only
freedom we have is in factor ordering and this was deemed
insufficient to alter the status-quo provided by the \WDW theory.

The situation changed dramatically in LQG. In contrast to the \WDW
theory, a well established, rigorous kinematical framework \emph{is}
available in full LQG \cite{almmt,alrev,crbook,ttbook}. If one
mimics it in symmetry reduced models, one is led to a quantum theory
which is \emph{inequivalent to the \WDW theory already at the
kinematic level}. Quantum dynamics built in this new arena agrees
with the \WDW theory in `tame' situations but differs dramatically
in the Planck regime, leading to a natural resolution of the big
bang singularity.

These developments occurred in three stages, each of which involved
major steps that overcame limitations of the previous one. As a
consequence, the viewpoint and the level of technical discussions
has evolved quite a bit. Therefore many statements made in the
literature have become outdated. Since non-experts can be confused
by the occasional tension between statements made at different
stages of this evolution, I will now make a small detour to
summarize how the subject evolved.

\subsection{Development of the subject}
 \label{s4.1}

The first seminal contribution was Bojowald's result \cite{mb1} that
the quantum Hamiltonian constraint of LQC does not break down at
$a=0$ where the classical singularity occurs. Since this was a major
shift that overcame the perceived impasse suggested by the \WDW
theory, it naturally led to a flurry of activity and the subject
began to develop. This success naturally drew scrutiny. Soon it
became clear that these fascinating results came at a cost: it was
implicitly assumed that $K$, the trace of the extrinsic curvature
(or the Hubble parameter, $\dot{a}/a$), is periodic, i.e. takes
values on a circle rather than the real line. Since this assumption
has no physical basis, at a 2002 workshop at Schr\"odinger
Institute, doubts arose as to whether the unexpectedly good behavior
of the quantum Hamiltonian constraint was an artifact of this
assumption.

However, thanks to key input from Klaus Fredenhagen at the same
workshop, it was soon realized \cite{abl} that in cosmological
models one can mimic the procedure used in full LQG to remove the
periodicity assumption. In the full theory, the requirement of
diffeomorphism covariance leads one to a \emph{unique}
representation of the algebra of fundamental operators \cite{lost}.
If one mimics in LQC the procedure followed in the full theory, one
finds $K$ naturally takes values on the real line as one would want
physically. But as I mentioned above, the resulting quantum
kinematics is \emph{inequivalent} to that of the \WDW theory and the
quantum Hamiltonian constraint is now a regular operator without
having to assume periodicity in $K$. This new kinematical framework
ushered-in the second stage of LQC. A number of early papers based
on periodicity of $k$ cannot be taken at their face value but
results of \cite{abl} suggested how they could be reworked in the
new kinematical framework. This led to another flurry of activity in
which more general models were considered. However, at this stage,
none of the analyses had a physical Hilbert space nor well-defined
Dirac observables. Indeed, often the Hamiltonian constraint failed
to be self-adjoint on the kinematical Hilbert space whence one could
not even begin to use the group averaging method. Consequently, new
questions arose. In particular, Brunnemann and Thiemann \cite{bt}
were led to ask: What is the precise sense in which the physical
singularity is resolved?

To address these key physical questions, one needs a physical
Hilbert space and a complete family of Dirac observables at least
some of which diverge at the singularity in the classical theory.
Examples are matter density, anisotropic shears and curvature
invariants (all evaluated at an instant of a suitably chosen
internal time). The question then is: Do the corresponding
operators all remain bounded on the \emph{physical} Hilbert space
even in the deep Planck regime? If so, one can say that the
singularity is resolved in the quantum theory. In the \WDW theory,
for example, these observables fail to remain bounded whence the
singularity is not resolved. What is the situation in LQC?

The third stage of evolution of LQC began with the detailed
construction of a mathematical framework to address this issue
\cite{aps1,aps2,aps3}. The physical Hilbert space could again be
constructed using the massless scalar field $\phi$ as internal time.
It was found \cite{aps2} that the self-adjoint version of the
Hamiltonian constraint introduced in the second stage \cite{abl}
---called the $\mu_o$ scheme in the literature--- does lead to
singularity resolution in the precise sense mentioned above. Since
the detailed theory could be constructed, the Hamiltonian constraint
could be solved numerically to extract physics in the Planck regime.

But this detailed analysis also brought out glaring limitations of
the theory which had remained unnoticed because the physical sector
of the theory had not been constructed. (For details see, e.g.,
Appendix 2 of \cite{aps3}, and \cite{cs}.) First, the matter density
(or curvature scalar) at which the bounce occurs depended
sensitively on quantum states and, unfortunately, more semiclassical
the state, lower is the density at which the bounce occurred.
Therefore the theory predicted huge deviations from general
relativity even at the density of water! Second, when cosmological
constant is non-zero, large deviations from general relativity can
occur even at late times, i.e., \emph{well away} from the Planck
regime. This is a concrete manifestation of the ultraviolet-infrared
tension discussed in section \ref{s2}. Finally, this quantum
dynamics has a deep conceptual flaw: in the k=0 model, even the
leading order predictions depend on the choice of the fiducial cell,
an auxiliary structure which has no physical significance. This
meant that, in spite of the singularity resolution, the theory was
physically unacceptable.

Fortunately, the problem could be traced back to the fact that
quantization \cite{abl,aps2} of the Hamiltonian constraint had
ignored a conceptual subtlety. Roughly, at a key step in the
procedure, the Hamiltonian constraint operator of \cite{abl}
implicitly used the kinematic metric $q^o_{ab}$ rather than the
physical metric $q_{ab}$. When this is corrected, the new, improved
Hamiltonian constraint again \emph{resolves the singularity and, at
the same time, is free from all the drawbacks of the $\mu_o$
scheme}. This is one of the best examples of the deep interplay
between physics and mathematics that I have encountered. The
improved procedure is referred to as the `$\bar\mu$ scheme' in the
literature. The resulting quantum dynamics has been analyzed in
detail and has provided a number of insights on the nature of
physics in the Planck regime (see section \ref{s4.2}). $\bar\mu$
dynamics has been successfully implemented in the case of a non-zero
cosmological constant \cite{aps3,bp}, the k=1, spatially compact
case \cite{apsv,warsaw1}, and to the Bianchi I model \cite{awe2}. As
I mentioned earlier, in the k=1 model, Green and Unruh \cite{gu} had
laid out more stringent tests that LQC has to meet to ensure that it
has good infrared behavior. These were met successfully. Because of
these advances, the $\bar\mu$ strategy has received considerable
attention from a mathematical physics perspective
\cite{warsaw1,warsaw2}. This work uses a combination of analytic and
numerical techniques to enhance rigor to a level that is
unprecedented in quantum cosmology.

We are now in the fourth stage of LQC where two directions are being
pursued. In the first, emphasis is on extending the framework to
more and more general situations (see in particular
\cite{aps4,hybrid}). Already in the spatially homogeneous
situations, the transition from $\mu_o$ to $\bar\mu$ scheme taught
us that great care is needed in the construction of the quantum
Hamiltonian constraint. The analysis of Bianchi I models has
re-enforced my belief that this lesson is an extremely valuable
guide for generalizations. It narrows down choices by making direct
appeal to physical considerations. The second direction in current
work is LQC phenomenology. Various LQC effects are being
incorporated in the analysis of observed properties of CMB
particularly by cosmologists (see, e.g.,
\cite{lqcp1,lqcp2,lqcp3,lqcp4,lqcp5,lqcp6}). These investigations
explore a wide range of issues, including: i) effects of the
quantum-geometry driven super-inflation just after the big-bounce,
predicted by LQC; ii) LQC driven effects on dark energy and phantom
fields;  and, iii) production of gravitational waves near the big
bounce. They combine very diverse ideas and are therefore important.
However, since most of this analysis has been carried out by
cosmologists who are not experts in quantum gravity, it does not
always do justice to all the relevant features, predictions and
subtleties of LQC. This frontier is still to mature. But the body of
work that has accumulated so far does provide an excellent
scaffolding and LQC community has begun to use it to build a more
careful, reliable and detailed framework.

\subsection{New quantum mechanics}
\label{s4.2}

Let us return to the difference between the \WDW theory and LQC. As
I mentioned at the beginning of section \ref{s4}, in the eighties
and nineties von-Neumann's uniqueness theorem and results from the
\WDW theory had led to a general belief that the big bang
singularity can not be resolved in quantum cosmology. So what
happens to the von Neumann theorem in LQC?

Let us first recall the statement of the theorem: 1-parameter
groups $U(\lambda)$ and $V(\mu)$ satisfying the Weyl commutation
relations%
\footnote{These are: $U(\lambda) V(\mu) = e^{i(\lambda\mu/\hbar)}
\,\,V(\mu) U(\lambda)$ and can be obtained by setting $U(\lambda)
= e^{i\lambda \h{x}}$ and $V(\mu) = e^{i\mu \h{p}}$ in the
standard Schr\"odinger theory. Given a representation,
$U(\lambda)$ is said to be \emph{weakly continuous} in $\lambda$
if its matrix elements between any two fixed quantum states are
continuous in $\lambda$.}
admit (up to isomorphism) a unique irreducible representation by
unitary operators on a Hilbert space $\H$ in which $U(\lambda)$ and
$V(\mu)$ are weakly continuous in the parameters $\lambda$ and
$\mu$. By Stone's theorem, weak continuity is a necessary and
sufficient condition for $\H$ to admit self adjoint operators
$\h{x}, \h{p}$ such that $U(\lambda) = e^{i\lambda \h{x}}$ and
$V(\mu) = e^{i\mu \h{p}}$. Therefore assumption of the von Neumann
theorem are natural in non-relativistic quantum mechanics and we are
led to a unique quantum kinematics ---the Schr\"odinger theory.

However, in full loop quantum gravity, $x$ is analogous to the
gravitational connection and $U(\lambda)$ to its holonomy. One can
again construct an abstract algebra using holonomies and operators
conjugate to connections and ask for its representations satisfying
natural assumptions, the most important of which is the
diffeomorphism covariance dictated by background independence. There
is again a uniqueness theorem \cite{lost}. However, in the
representation that is thus singled out, holonomy operators
---analogs of $U(\lambda)$--- \emph{fail to be weakly continuous}
whence there are no operators corresponding to connections!
Furthermore, a number of key features of the theory
---such as the emergence of a quantum Riemannian geometry in which
there is fundamental discreteness--- can be traced back to this
unforeseen feature. Therefore, upon symmetry reduction, although we
have a finite number of degrees of freedom, it would be incorrect to
just mimic Schr\"odinger quantum mechanics and impose weak
continuity. When this assumption is dropped, the von Neumann theorem
is no longer applicable and \emph{we can have new quantum mechanics}
\cite{abl}. This new kinematical arena is constructed by applying
the procedure used in full LQG to the symmetry reduced models of
LQC.

Thus, the key difference between LQC and the \WDW theory lies in the
fact that while one does not have reliable quantum kinematics in the
\WDW theory, there is a well developed and rigorous framework in LQG
which, furthermore, is \emph{unique}! If we mimic it as closely as
possible in the symmetry reduced theories, we are led to a new
kinematic arena, distinct from the one used in the \WDW quantum
cosmology. LQC is based on this arena.

\section{LQC: Dynamics}
 \label{s5}

It turns out \WDW dynamics is not supported by the new kinematical
arena because, when translated in terms of gravitational connections
and their conjugate momenta, it requires that there be an operator
corresponding to the connection itself. As we just saw, such as
operator does not exist on the kinematical Hilbert space $\Hk$ of
LQC; only its exponentiated versions, the holonomies, are well
defined on $\Hk$. Therefore one has to develop quantum dynamics
ab-initio on the new arena (see appendix \ref{a2.1}).

Now in the Hamiltonian constraint of LQG, the gravitational spin
connection $A_a^i$ appears through its curvature $F_{ab}{}^i$. Since
only holonomy operators are well-defined, we are led to express
curvature in terms of holonomies. In the classical theory, various
components of $F_{ab}{}^i$ can be obtained by first computing
holonomies around appropriate (and naturally available) loops,
dividing them by the areas enclosed by these loops, and taking the
limit as the area shrinks to zero. In the quantum theory, the limit
does not exist, reflecting that there is no local operator
representing $A_a^i$. This, as we saw, is a distinguishing feature
of LQG which lies at the heart of the discreteness of geometric
operators. In particular, there is a minimum non-zero area
eigenvalue ---$\Delta := 4\sqrt{3}\pi\gamma\lp^2$ in Planck units,
where $\gamma$ is the Barbero-Immirzi parameter--- of the area
operator, often referred to as the area gap. The fact that the
fundamental quantum geometry has a built-in discreteness is a strong
hint that we should shrink the loop only till the \emph{physical}
area it encloses is $\Delta\lp^2$.%
\footnote{In the $\mu_o$ framework, this area was computed using the
fiducial, rather than the physical, metric (i.e., $q^o_{ab}$ rather
than $q_{ab}$). As we noted above the resulting quantum dynamics has
a good ultra-violet behavior but a bad infra-red behavior. This is
cured in the $\bar\mu$ scheme simply by using the physical metric to
compute these areas. Secondly, in the $\bar\mu$ scheme,
$\Delta\lp^2$ was at first taken to be the lowest non-zero
eigenvalue $2\sqrt{3}\pi \gamma \ell_{\rm{Pl}}^2$
\cite{aps1,aps2,aps3,acs} (where $\gamma$ is the Barbero-Immirzi
parameter). However, it was later realized \cite{awe2} that
eigenstates with these eigenvalues will not appear in quantum
geometries needed to represent homogeneous classical metrics. On
states that can feature in this representation, the minimum non-zero
eigenvalue is twice as large. Since the critical density $\rcr$ at
which the quantum bounce occurs in these models goes inversely as
$\Delta$, in the earlier literature, $\rcr$ was $\sim
0.82\rho_{\rm{Pl}}$  rather than $\sim 0.41\rho_{\rm Pl}$. Both thes
points are discussed further in appendix \ref{a2.1}.}
As a consequence, the quantum curvature operator $\h{F}_{ab}{}^i$
---and hence the quantum dynamics--- is now \emph{non-local}.
Locality is recovered only in the classical limit.

These considerations lead to a well-defined Hamiltonian constraint
operator on the kinematical Hilbert space of LQC. It has the same
form $\partial_\phi^2 \Psi(v, \phi) = - \Theta \Psi(v,\phi)$ as in
the \WDW theory but differential operator $\ul\Theta_o\, =\, -12\pi
G\, (v\partial_v)^2\,$ that features in the \WDW constraints Eqs
(\ref{wdw0}) and (\ref{wdw1}) is now replaced by a second order
\emph{difference} operator $\Theta_o$ in $v$:
\be \label{hc4} \Theta_o \Psi(v,\phi) =  - F(v) \, \left(C^+(v)\,
\Psi(v+4,\phi) + C^o(v) \, \Psi(v,\phi) +C^-(v)\,
\Psi(v-4,\phi)\right)\, . \ee
Here, $F(v)$ and $C^\pm(v)$ and $C^o(v)$ are functions of $v$:
\be \label{B} F(v) = \f{3\sqrt{3\sqrt{3}}}{2\sqrt{2}}\,\, |v| \, .
\ee
and
\ba \label{C} C^+(v) &=& \nonumber \f{3\pi K G}{8} \, |v + 2| \,\,\,
\big| |v + 1| - |v +3|  \big|  \\
C^-(v) &=& \nonumber C^+(v - 4) \quad {\rm and} \quad C^o(v) = -
C^+(v) - C^-(v) ~. \ea
As one might expect, the step size in the difference operator
$\Theta_o$ is dictated by the area gap $\Delta \lp^2$. There is a
precise sense in which the \WDW equations emerge as limits of LQC
equations when $\Delta$ is taken to zero, i.e., when the Planck
scale discreteness of quantum geometry determined by LQG is
neglected (see appendix \ref{a2.4}). Consequently, discreteness in
LQC dynamics is completely negligible at late times. However, as we
will see, it plays a crucial role in the Planck scale regime near
singularities.

Dynamics dictated by this difference equation has been analyzed
using three different methods:
\smallskip

\noindent \b Numerical solutions of the exact quantum equations
    \cite{aps1,aps2,aps3,apsv}. A great deal of effort was spent in
    ensuring that the results are free of artifacts of simulations,
    do not depend on the details of how semi-classical states are
    constructed and hold for a wide range of
    parameters.\\
\b Effective equations \cite{jw,aps3,apsv,cv}. These are
    differential equations which include the leading quantum
    corrections. The asymptotic series from which these
    contributions were picked was constructed rigorously but is
    based on assumptions whose validity has not been
    established. Nonetheless the effective equations approximate
    the exact numerical evolution of semi-classical states
    extremely well.\\
\b Exact analytical results in the k=0, $\Lambda$=0 model
    \cite{mb-exact,acs}. This analysis has provided an
    analytical understanding of some of the numerical results as
    well as several new results which are not restricted to
    states that are semi-classical at late times \cite{acs}. In
    this sense the overall picture is robust within these
    models.\\

I will now provide a global picture that has emerged from these
investigations, first for the k=1 model without the cosmological
constant $\Lambda$ and for the k=0 case for various values of
$\Lambda$. Recall that in classical general relativity, the k=1
closed universes start out with a big bang, expand to a maximum
volume $V_{\rm max}$ and then recollapse to a big-crunch
singularity. Consider a classical solution in which $V_{\rm max}$ is
astronomically large ---i.e., on which the constant of motion
$p_{(\phi)}$ takes a large value $p_{(\phi)}^\star$--- and consider
a time $\phi_o$ at which the volume $v^\star$ of the universe is
also large. Then there are well-defined procedures to construct
states $\Psi(v,\phi)$ in the \emph{physical Hilbert space} which are
sharply peaked at these values of observables $\h{p}_{(\phi)}$ and
$\h{V}|_{\phi_o}$ at the `time' $\phi_o$. Thus, at `time' $\phi_o$,
the quantum universe is well approximated by the classical one. What
happens to such quantum states under evolution? As emphasized
earlier, there are infra-red and ultra-violet challenges:\\
i) Does the state remain peaked on the classical trajectory in the
low curvature regime? Or, do quantum geometry effects accumulate
over the cosmological time scales, causing noticeable deviations
from classical general relativity? In particular, as Green and Unruh
\cite{gu} asked, is there a recollapse and if so does the value
$V_{\rm max}$ of maximum volume agree with that predicted by general
relativity?\\
ii) What is the behavior of the quantum state in the Planck regime?
Is the big-bang singularity resolved? What about the big-crunch? If
they are both resolved, what is on the `other side'?

\begin{figure}[]
  \begin{center}
    $a)$\hspace{8cm}$b)$
    \includegraphics[width=3.2in,angle=0]{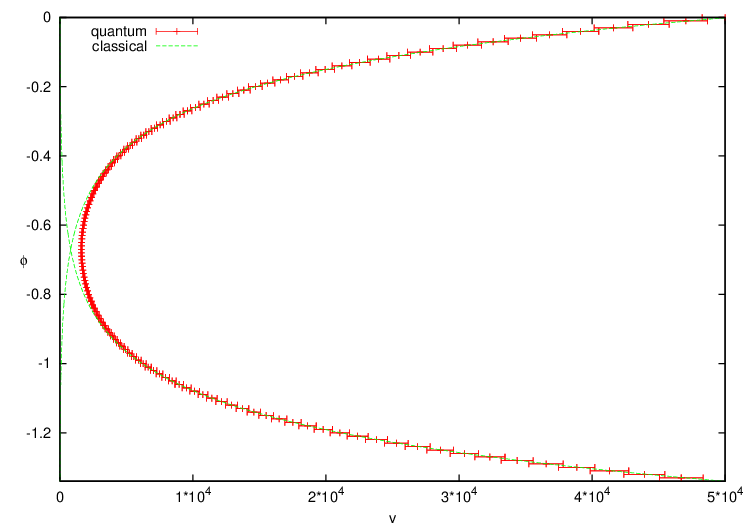}
    \includegraphics[width=3.2in,angle=0]{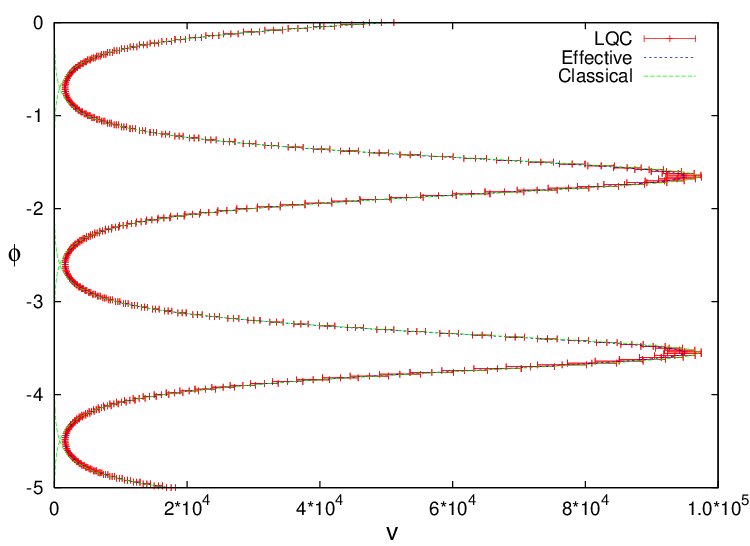}
\caption{In the LQC evolution of models under consideration, the big
bang and big crunch singularities are replaced by quantum bounces.
Expectation values and dispersion of $|\h{v}|_\phi$, are compared
with the classical trajectory. The classical trajectory deviates
significantly from the quantum evolution at the Planck scale and evolves
into singularities. By contrast, the effective trajectory provides
an excellent approximation to the quantum evolution at all scales.
\,\, $a)$ The k=0 case. In the backward evolution, the quantum
evolution follows our post big-bang branch at low densities and
curvatures but undergoes a quantum bounce at matter density $\rho
\sim 0.41\rho_{\rm PL}$ and joins on to the classical trajectory
that was contracting to the future. $b)$ The k=1 case. The quantum
bounce occurs again at $\rho \sim 0.41 \rho_{\rm Pl}$. Since the big
bang and the big crunch singularities are resolved the evolution
undergoes cycles. In this simulation $p_{(\phi)}^\star = 5\times
10^3$, $\Delta p_{(\phi)}/p_{(\phi)}^\star = 0.018$, and $v^\star =
5\times 10^4$.} \label{fig:lqc}
\end{center}
\end{figure}

Numerical simulations show that the wave functions do remain sharply
peaked on classical trajectories in the low curvature region also in
LQC. But there is a radical departure from the \WDW results in the
strong curvature region: While the \WDW evolution follows classical
dynamics all the way into the big-bang and big crunch singularities
(see Fig \ref{fig:wdw}), in LQC \emph{the big bang and the big
crunch singularities are resolved and replaced by big-bounces} (see
Fig \ref{fig:lqc}). In these calculations, the required notion of
semi-classicality turns out to be surprisingly weak: these results
hold even for universes with $a_{\rm max} \approx 23 \lp$ and the
`sharply peaked' property improves greatly as $a_{\mathrm{max}}$
grows. More precisely, numerical solutions have shown that the
situation is as follows. (For details, see \cite{apsv}.)
\smallskip

\b The trajectory defined by the expectation values of the physical
observable $\h{V}|_\phi$ in the full quantum theory is in good
agreement with the trajectory defined by the classical Friedmann
dynamics until the energy density $\rho$ in the matter field is
about one percent of the Planck density. In the classical solution,
scalar curvature and the matter energy density keep increasing on
further evolution, eventually leading to a big bang (respectively,
big crunch) singularity in the backward (respectively, forward)
evolution, where $v \rightarrow 0$. The situation is very different
with quantum evolution. As the density increases beyond $0.01
\rho_{\rm Pl}$, quantum geometry effects become dominant, creating
an effective repulsive force which rises very quickly. It then
overwhelms classical gravitational attraction, and causes a bounce
when $\rho$ reaches a critical value $\rcr \sim 0.41\rho_{\rm Pl}$,
thereby resolving the past (or the big bang) and future (or the big
crunch) singularities. Thus, there is a cyclic scenario depicted in
Fig \ref{fig:lqc}.

\b The volume of the universe takes its minimum value $V_{\rm min}$
at the bounce point. $V_{\rm min}$ scales linearly with
$p_{(\phi)}$:
\be V_{\rm min}\,\, \propto \,\, p_{(\phi)} \ee
%
Consequently, $V_{\rm min}$ can be \emph{much} larger than the
Planck size.  Consider for example a quantum state describing a
universe which attains a maximum radius of a megaparsec. Then the
quantum bounce occurs when the volume reaches the value $V_{\rm min}
\approx 5.7 \times 10^{16}\, {\cm}^3$, \emph{some $10^{115}$ times
the Planck volume.} Deviations from the classical behavior are
triggered when the density or curvature reaches the Planck scale.
The volume can be very large; it is not the variable that sets the
relevant scale for quantum gravity effects.

\b After the quantum bounce the energy density of the universe
decreases and the repulsive force dies quickly when matter density
reduces to about two percept of the Planck density. The quantum
evolution is then well-approximated by the classical trajectory. On
subsequent evolution, the universe recollapses both in classical and
quantum theory at the value $V=V_{\mathrm{max}}$ when energy density
reaches a minimum value $\rmin$.  $V_{\rm max}$ scales as the
3/2-power of $p_{(\phi)}$:
\be\label{Vmax}  V_{\rm max}  
\,\, \propto \,\,  p_{(\phi)}^{{3/2}} \ee
Quantum corrections to the classical Friedmann formula $\rmin =
3/8\pi Ga^2_{\rm max}$ are of the order $O(\lp/a_{\rm max})^4$. For
a universe with $a_{\rm max} = 23\lp$, the correction is only one
part in $10^{5}$. For universes which grow to macroscopic sizes,
classical general relativity is essentially exact near the
recollapse.

\b Using ideas from geometrical quantum mechanics \cite{as}, one can
obtain certain effective classical equations which incorporate the
leading quantum corrections \cite{jw,apsv}. As one would expect,
quantum geometry effects primarily modify the left hand side of
Einstein's equations. However, in the simplest models considered
here, it is possible to rewrite the quantum corrected equation by
moving the correction to the right side and then expressing it in
terms of matter. This rewriting is convenient in comparing the
effective equations with the classical ones. The classical Friedmann
equation, $(\dot{a}/a)^2 = ({8\pi G}/{3})\, (\rho- 3/8\pi G a^2)$,
is replaced by
\be \label{eff} \left(\f{\dot{a}}{a} \right)^2 = \fs{8\pi G}{3}\,
(\rho-\rho_1(v))\,\, \left[f(v) - \f{\rho}{\rcr} \right]\, , \ee
where $\rho_1$ and $f$ are specific functions of $v$ with $\rho_1
\sim 3/8\pi G a^2$ for large $a$. Bounces occur when $\dot{a}$
vanishes, i.e. at the value of $v$ at which the matter density
equals $\rho_1(v)$ or $\rcr\,f(v)$. The first root $\rho(v) =
\rho_1(v)$ corresponds to the classical recollapse while the second
root, $\rho=\rcr\,f(v)$, to the quantum bounce. Away from the Planck
regime, $f \approx 1$ and $\rho/\rho_{\rm crit} \approx 0$.

\b For quantum states under discussion, the density $\rmax$ is well
approximated by $\rcr \approx 0.41 \rho_{\mathrm{Pl}}$ up to terms
$O(\lp^2/a_{\rm min}^2)$, independently of the details of the state
and values of $p_{(\phi)}$. (For a universe with maximum radius of a
megaparsec, $\lp^2/a_{\rm min}^2 \approx 10^{-76}$.) The density
$\rmin$ at the recollapse point also agrees with the value $(3/8 \pi
G a^2_{\rm max})$ predicted by the classical evolution to terms of
the order $O(\lp^4/a_{\rm max}^4)$. Furthermore the scale factor
$a_{\rm max}$ at which recollapse occurs in the quantum theory
agrees to a very good precision with the one predicted by the
classical dynamics.

\b The trajectory obtained from effective Friedmann dynamics is in
excellent agreement with quantum dynamics \emph{throughout the
evolution.}  In particular, the maximum and the minimum energy
densities predicted by the effective description agree with the
corresponding expectation values of the density operator $\hat \rho
\equiv \widehat{p_{(\phi)}^2/2|p|^3}$ computed numerically.

\b The state remains sharply peaked for a \emph{very large number of
`cycles'.} Consider the example of a semi-classical state with an
almost equal relative dispersion in $p_{(\phi)}$ and $|v|_\phi$ and
peaked at a large classical universe of the size of a megaparsec.
When evolved, it remains sharply peaked with relative dispersion in
$|v|_\phi$ of the order of $10^{-6}$ \emph{even after $10^{50}$
cycles of contraction and expansion!} Any given quantum state
eventually ceases to be sharply peaked in $|v|_\phi$ (although it
continues to be sharply peaked in the constant of motion
$p_{(\phi)}$). Nonetheless, the quantum evolution continues to be
deterministic and well-defined for an infinite number of cycles.
This is in sharp contrast with the classical theory where the
equations break down at singularities and there is
no deterministic evolution from one cycle to the next. \\

This concludes the summary of our discussion of the k=1 model.  An
analogous detailed analysis has been carried out also in the k=0
model, again with a free massless scalar field
\cite{aps1,aps2,aps3,acs}. In this case, if the cosmological
constant $\Lambda$ vanishes, as Fig \ref{class} shows, classical
solutions are of two types, those which start out at the big-bang
and expand out to infinity and those which start out with large
volume and contract to the big crunch singularity. Again, in this
case while the \WDW solution follows classical trajectories into
singularities, the LQC solutions exhibit a big bounce. The LQC
dynamics is again faithfully reproduced by an effective equation.
Again, quantum geometry modifies the left hand side of Einstein's
equations but one can move this correction to the right side through
an algebraic manipulation. Then, one finds that the Friedmann
equation $(\dot{a}/a)^2 = (8\pi G\, \rho/3)$ is replaced by
\be \left( \f{\dot{a}}{a} \right)^2 = (8\pi G\,\rho /3)\, \left(1 -
\f{\rho}{\rcr}\right) \, . \ee
In classical general relativity, the right side,\, $8\pi G
\rho/3$,\, is positive, whence $\dot{a}$ cannot vanish; the universe
either expands forever from the big bang or contracts into the big
crunch. In the LQC effective equation on the other hand, $\dot{a}$
vanishes when $\rho=\rcr$ at which a quantum bounce occurs: To the
past of this event, the universe is contracts while to the future,
it expands. This is possible because the LQC correction $\rho/\rcr$
\emph{naturally} comes with a negative sign. This is non-trivial. In
the standard brane world scenario, for example, Friedmann equation
is also receives a $\rho/\rcr$ correction but it it comes with a
positive sign (unless one artificially makes the brane tension
negative) whence the singularity is not resolved.

Even at the onset of the standard inflationary era, the quantum
correction $\rho/\rcr$ is of the order $10^{-11}$ and hence
completely negligible. Thus, we are justified in using classical
general relativity during inflation. The quantum bounce occurs at
$\rho=\rcr$ and the critical density is again given by $\rcr \approx
0.41\rho_{\rm Pl}$. Furthermore, one can show \cite{acs}
analytically that the spectrum of the density operator \emph{on the
physical Hilbert space} admits a finite upper bound $\rho_{\rm
sup}$,
\be \rho_{\rm sup} = \f{\sqrt{3}}{32\pi^2\gamma^3G^2\hbar}\,\, . \ee
By plugging values of constants in the analytical expression of this
bound, one finds $\rho_{\rm sup} = \rcr$!%
\footnote{In this evaluation, one uses the value $\gamma \approx
0.2375$ of the Barbero-Immirzi parameter $\gamma$ obtained from
black hole entropy calculations. The numerical simulations used to
calculate $\rcr$ use the same value. Because of the factor
$\gamma^{-3}$, the value of $\rho_{\rm sup}$, is quite sensitive to
that of $\gamma$. The fact that $\rho_{\rm sup}$ is of the order of
$\rho_{\rm Pl}$ brings out a pleasing coherence between LQC and the
entropy calculation from LQG.}

Inclusion of a cosmological constant is discussed in \cite{bp}. If
$\Lambda >0$, there are again two types of classical trajectories
but the one which starts out at the big-bang expands to an infinite
volume at a \emph{finite} value $\phi_{\rm max}$ of $\phi$. The
energy density $\rho_\phi$ in the scalar field goes to zero at
$\phi_{\rm max}$. (The other trajectory is a `time reverse' of
this.) Because the $\phi$ `evolution' is unitary in LQC, it yields a
natural extension of the classical solution beyond $\phi_{\rm max}$.
States which are semi-classical in the low $\rho_\phi$ regime again
follow an effective trajectory. Since $\rho_\phi$ remains bounded,
it is convenient to draw these trajectories in the
$\rho_\phi$-$\phi$ plane (rather than $v$-$\phi$ plane). They agree
with the classical trajectories in the low $\rho_\phi$ regime and
analytically continue the classical trajectories beyond $\rho_{\phi}
=0$. If $\Lambda <0$, the classical universe undergoes a recollapse.
This is faithfully reproduced by the LQC evolution. Since both the
big-bang and the big-crunch singularities are resolved, the LQC
evolution leads to a cyclic universe as in the k=1 model. Thus, in
all these cases, the principal features of the LQC evolution are
robust, including the value of $\rcr$.

\section{Discussion}
\label{s6}

Let us summarize the overall situation. In simple cosmological
models, all the questions raised in section \ref{s2} have been
answered in LQC in remarkable detail. The scalar field plays the
role of an internal or emergent time and enables us to interpret the
Hamiltonian constraint as an evolution equation. The matter momentum
$\h{p}_{(\phi)}$ and `instantaneous' volumes $\h{V}|_\phi$ form a
complete set of Dirac observables. They enable us to ask physically
interesting questions. In particular, the density $\rho$ (and the
4-d Ricci scalar) always remains bounded. In this precise sense, the
big bang and the big crunch singularities are naturally resolved. On
the `other side' of the bounce there is again a large universe.
General relativity is an excellent approximation to quantum dynamics
once the matter density falls below a percent of the Planck density.
Thus, LQC successfully meets both the `ultra-violet' and `infra-red'
challenges. Furthermore results obtained in a number of models using
distinct methods re-enforce one another. One is therefore led to
take at least the qualitative findings seriously: \emph{Big bang is
not the Beginning nor the big crunch the End.} Quantum space-time
appears to be vastly larger than what general relativity had us
believe!

In LQC, main departures from the \WDW theory occur due to
\emph{quantum geometry effects} of LQG. There is no fine tuning of
initial conditions, nor a boundary condition at the singularity,
postulated from outside. Also, there is no violation of energy
conditions. In fact quantum corrections to the matter Hamiltonian do
not play any role in the resolution of singularities of these
models. John Wheeler's vision, summarized in the quote that I began
this article with is realized to a surprising extent: Indeed,
\begin{quote}
{\sl In a finite proper time the calculated curvature rises to
infinity} [in the classical theory]. {\sl At this point the
classical theory becomes incapable of further prediction. In
actuality, classical predictions go wrong before this point. ... The
semiclassical treatment of propagation is appropriate in most of the
domain of superspace of interest to gravitational collapse. Not so
in the decisive region.}
\end{quote}
Wheeler was very fond of the fundamental discreteness underlying
loop quantum gravity. So, he would have been pleased that it is
precisely this feature that is responsible for the singularity
resolution.

If no energy conditions are violated, how does this singularity
resolution square with the general singularity theorems of Penrose
and Hawking? They are evaded because \emph{the left hand side} of
the classical Einstein's equations is modified by the quantum
geometry corrections of LQC. What about the more recent singularity
theorems that Borde, Guth and Vilenkin \cite{bgv} proved in the
context of inflation? They do not refer to Einstein's equations.
But, motivated by the eternal inflationary scenario, they assume
that the expansion is positive along any past geodesic. Because of
the pre-big-bang contracting phase, this assumption is violated in
the LQC effective theory.

While the detailed results presented in section \ref{s5} are valid
only for these simplest models, partial results have been obtained
also in more complicated models indicating that the singularity
resolution may be robust \cite{hybrid} (in the sense of geodesic
completeness \cite{ps}). In this respect there is a curious
similarity with the very discovery of singularities in general
relativity. They were first encountered in special examples.
Although the examples were the physically most interesting ones
---e.g., the big-bang and the Schwarzschild curvature
singularities--- at first it was thought that these space-times are
singular because they are highly symmetric. It was widely believed
that generic solutions of Einstein'e equations should be
non-singular. As is well-known, this belief was shattered by the
singularity theorems. Some 40 years later we have come to see that
the big bang and the big crunch singularities are in fact resolved
by quantum geometry effects. Is this an artifact of high symmetry?
Or, are there robust \emph{singularity resolution theorems} lurking
just around the corner \cite{ps}?

A qualitative picture that emerges is that the non-perturbative
quantum geometry corrections are \emph{`repulsive'}.%
\footnote{We saw in section \ref{s4} that there is no connection
operator in LQG. As a result the curvature operator has to be
expressed in terms of holonomies and becomes non-local. The
repulsive force can be traced back to this non-locality.
Heuristically, the polymer excitations of geometry do not like to
be packed too densely; if brought too close, they repel.}
While they are negligible under normal conditions, they dominate
when curvature approaches the Planck scale and can halt the collapse
that would classically have led to a singularity. In this respect,
there is a curious similarity with the situation in the stellar
collapse where a new repulsive force comes into play when the core
approaches a critical density, halting further collapse and leading
to stable white dwarfs and neutron stars. This force, with its
origin in the Fermi-Dirac statistics, is \emph{associated with the
quantum nature of matter}. However, if the total mass of the star is
larger than, say, $5$ solar masses, classical gravity overwhelms
this force. The suggestion from LQC is that a new repulsive force
\emph{associated with the quantum nature of geometry} comes into
play and is strong enough to counter the classical, gravitational
attraction, irrespective of how large the mass is. It is this force
that prevents the formation of singularities. Since it is negligible
until one enters the Planck regime, predictions of classical
relativity on the formation of trapped surfaces, dynamical and
isolated horizons would still hold. But one cannot conclude that
there must be a singularity because the assumptions of the standard
singularity theorems would be violated. There may be no
singularities, no abrupt end to space-time where physics stops.
Non-perturbative, background independent quantum physics would
continue.

At first one might think that, since quantum gravity effects concern
only a tiny region, whatever quantum effects there may be, their
influence on the global properties of space-time should be
negligible whence they would have almost no bearing on the issue of
the Beginning and the End. However, as we saw, once the singularity
is resolved, vast new regions can appear on the `other side'
ushering in new possibilities that were totally unforeseen in the
realm of classical general relativity. This year we celebrate the
100th anniversary of Minkowski's celebrated paper that fused space
and time into a 4-dimensional space-time continuum. It is fitting
that concrete hints on what may eventually replace that continuum
have begun to appear on the horizon.

\bigskip

\textbf{Acknowledgments:} Much of this chapter (and all figures it
contains) is based on joint work with Alex Corichi, Tomasz
Pawlowski, Param Singh, Victor Taveras, and Kevin Vandersloot. I
have also benefited from comments, suggestions and probing questions
of many colleagues, especially Martin Bojowald, James Hartle, Jerzy
Lewandowski, Donald Marolf, Roger Penrose, Carlo Rovelli and
Madhavan Varadarajan This work was supported in part by the NSF
grant PHY0456913, the Alexander von Humboldt Foundation, the The
George A. and Margaret M. Downsbrough Endowment and the Eberly
research funds of Penn State.

\begin{appendix}

\section{General conceptual issues}
\label{a1}

This appendix is divided into three parts. In the first I address
the question of why LQC is physically interesting in spite of its
drastic symmetry reduction; in the second I discuss the issue of
extracting dynamics from the `frozen formalism' and in the third I
analyze Bousso's covariant entropy bound from the LQC perspective.

\subsection{Why is the field of quantum cosmology interesting?}
\label{a1.1}

The symmetry reduction used to descend to quantum cosmology is
drastic because it ignores all but finitely many degrees of freedom
of the gravitational matter fields. Furthermore, it is this
simplification that enables one to obtain detailed predictions in
the deep Planck regime. So, a natural question arises: Why can we
trust quantum cosmology? Why is it useful? More precisely, in our
case, will predictions of full LQG resemble anything like what LQC
predicts? This is an important question. I would like to give three
arguments which suggest that the answer may be in the affirmative.

First, consider the analogy of electrodynamics. Suppose,
hypothetically, that we had full QED but somehow did not have a good
description of the hydrogen atom. (Indeed, it is very difficult to
have a complete control on this bound state problem in the framework
of full QED!) Suppose that Dirac came along at this juncture and
said: let us first impose spherical symmetry, describe the proton
and electron as particles, and then quantize the system. In this
framework, all radiative modes of the electromagnetic field would be
frozen and we would have quantum mechanics; the Dirac theory of
hydrogen atom. One's first reaction would have been that the
simplification involved is so drastic that there is no reason to
expect this theory to captures the essential features of the
physical problem. Yet we know it does. Quantum cosmology may well be
the analog of the hydrogen atom in quantum gravity.

Second, recall what happened in classical general relativity.
Singularities were first discovered in highly symmetric models. The
general wisdom derived from the detailed analysis of the Russian
school led by Khalatnikov, Lifshitz and others was that these
singularities were artifacts of the high symmetry and a generic
solution of Einstein's equations with physically reasonable matter
would be singularity free. As noted in section \ref{s6}, singularity
theorems of Penrose, Hawking, Geroch and others shattered this
paradigm. We learned that lessons derived from symmetry reduced
models were in fact much more general than anyone would have
suspected. The work of the Madrid group \cite{hybrid} and more
recent analysis of effective equations by Singh \cite{ps} suggests
that the situation may be similar with singularity resolution of
LQC.

Finally, the Belinskii-Khalatnikov-Lifshitz (BKL) conjecture in
classical general relativity says that as one approaches space-like
singularities in general relativity, `spatial derivatives of basic
fields become sub-dominant relative to the time derivatives' and
dynamics at any spatial point is well approximated by that of
homogeneous models, Bianchi I dynamics playing a dominant role (see,
e.g., \cite{bkl}). By now there is considerable support for this
conjecture both from rigorous mathematical and numerical investigations.%
\footnote{More recently, the conjecture has been formulated in the
framework one uses in LQG \cite{ahs}.}
This provides some support for the idea that lessons on the quantum
nature of the big-bang (and big-crunch) singularities
---particularly in the Bianchi I model \cite{awe2}---
may be valid much more generally.

Of course none of these arguments shows conclusively that the
qualitative features of LQC will remain in tact in the full
theory. But they do suggest that one should not a priori dismiss
LQC as being too simple.

\subsection{How do you extract physics from the frozen formalism
of Bergmann and Dirac?} \label{a1.2}

In the 60s and 70s Bergmann and Komar \cite{kk} pointed out that if
one follows the canonical quantization procedure developed by
Bergmann and Dirac, one is led to a `frozen formalism': since
physical states are solutions to the quantum constraints, (Dirac)
observables must commute with the constraints and therefore appear
to be constants of motion. To directly extract the dynamical content
of the theory from this frozen formalism, one can adopt the
following program discussed in sections \ref{s4} and \ref{s5}:\\
i) Isolate, among all dynamical variables one that can represent
`relational time' with respect to which other physical variables
`evolve'.\\
ii) Introduce an appropriate scalar product on the space of
solutions to the quantum constraints. This would be the physical
Hilbert space $\Hp$.\\
iii) Introduce on $\Hp$, a complete set of self-adjoint operators
which have direct physical interpretation. These would be the Dirac
observables which evolve with respect to the relational time
variable.\\
iv) Construct states which are semi-classical at late times, i.e.,
sharply peaked at a point on a classical trajectory where the matter
density and curvatures are all too small for quantum gravity effects
to be significant. Evolve them. Do they remain peaked on the
classical trajectory in the weak curvature regime? What happens to
the Dirac observables in the deep Planck regime? Do they remain
bounded or can they diverge?  What is the physical interpretation of
the part of the state to the `past of the big-bang singularity'?

In LQC this program was completed as follows. As we saw in section
\ref{s5}, in the simplest models the quantum Hamiltonian constraint
takes the form
 \be \label{qha1}\partial_\phi^2 \Psi(v, \phi) = -\Theta\, \Psi(v,\phi) \ee
where $\Theta$ is a positive, self-adjoint \textit{difference}
operator, independent of $\phi$. As explained in the main text, the
form of this constraint suggests that we use $\phi$ as the internal
or relational time variable and $\hat{V}$ (or, alternatively, matter
density $\hat\rho$) as the physical variable which evolves with
respect to $\phi$. The scalar product on the space of solutions to
(\ref{qha1}) can be introduced using the group averaging technique.
Since (\ref{qha1}) has the same form as the Klein-Gordon equation on
a static space-time, the resulting scalar product is completely
analogous to that for the Klein Gordon field. Thus, we can regard
the physical Hilbert space $\Hp$ as consisting of `positive
frequency solutions' to (\ref{qha1}), i.e. solutions to its positive
square root
\be \label{qha2} -i \partial_\phi \Psi(v,\phi) = \sqrt{\Theta}\,
\Psi(v,\phi). \ee
This has the form of the Schr\"odinger equation with Hamiltonian
$\h{H}= \sqrt{\Theta}$. Therefore, the inner product on the
gravitational kinematic Hilbert space $\Hkg$ with respect to which
$\Theta$ is a positive, self adjoint operator is conserved in
internal time $\phi$ and provides us with the physical Hilbert space
$\Hp$.

Finally we can define Dirac observables. First, since
$\h{p}_{(\phi)}$ is a constant of motion, it maps physical states to
physical states; it is a natural Dirac observable. The second Dirac
observable is the volume operator $\h{V}|_{\phi_o}$ at any fixed
`instant' $\phi=\phi_o$:
\be [\hat{V}|_{\phi_o} \Psi](v,\phi) = e^{i\sqrt{\Theta}(\phi -
\phi_o)}\, \hat{V}\, \Psi(v, \phi_o)\ee
A more interesting physical observable is the density operator
$\h{\rho}|_{\phi_o}$ at the instant of time $\phi_o$, defined by
$\hat\rho|_{\phi_o} := (1/2) \widehat{V^{-1}}|_{\phi_o}\,
\h{p}_{(\phi)}^2 \, \widehat{V^{-1}}|_{\phi_o}$. In the classical
theory, this observable diverges as one approaches the singularity.
In striking contrast, it remains bounded on the physical Hilbert
space $\Hp$ \cite{acs}. Finally, as we saw in the main text, if we
start with a semi-classical state in the distant future and evolve
back toward the singularity using (\ref{qha2}), we obtain a state
which is semi-classical also on the `other side of the big-bang'.

Thus, in simple cosmological models, there is a natural avenue to
extract physics from the `frozen formalism'.

\subsection{Can LQC `save' the covariant entropy bound?}
\label{a1.3}

Motivated by the black hole entropy formula, several entropy bounds
have been proposed in the literature. The heuristic idea is the
following. The leading contribution to the black hole entropy is
given by (1/4)th of the area of its (isolated) horizon in Planck
units. Since black holes are the `densest' objects, one is then led
to idea that if we had a complete quantum gravity theory, the number
of states in any volume $V$ enclosed by a surface of area $A$ should
be bounded by the number of states of a black hole with a horizon of
area $A$, i.e. by $\exp {A/4\lp^2}$. However, this simple formula
quickly runs into difficulties. Several improvements have been
proposed. The most developed of these proposals is Bousso's
covariant entropy bound \cite{bousso1}.

\begin{figure}
\begin{center}
\includegraphics[width=1.8in,height=1.8in]{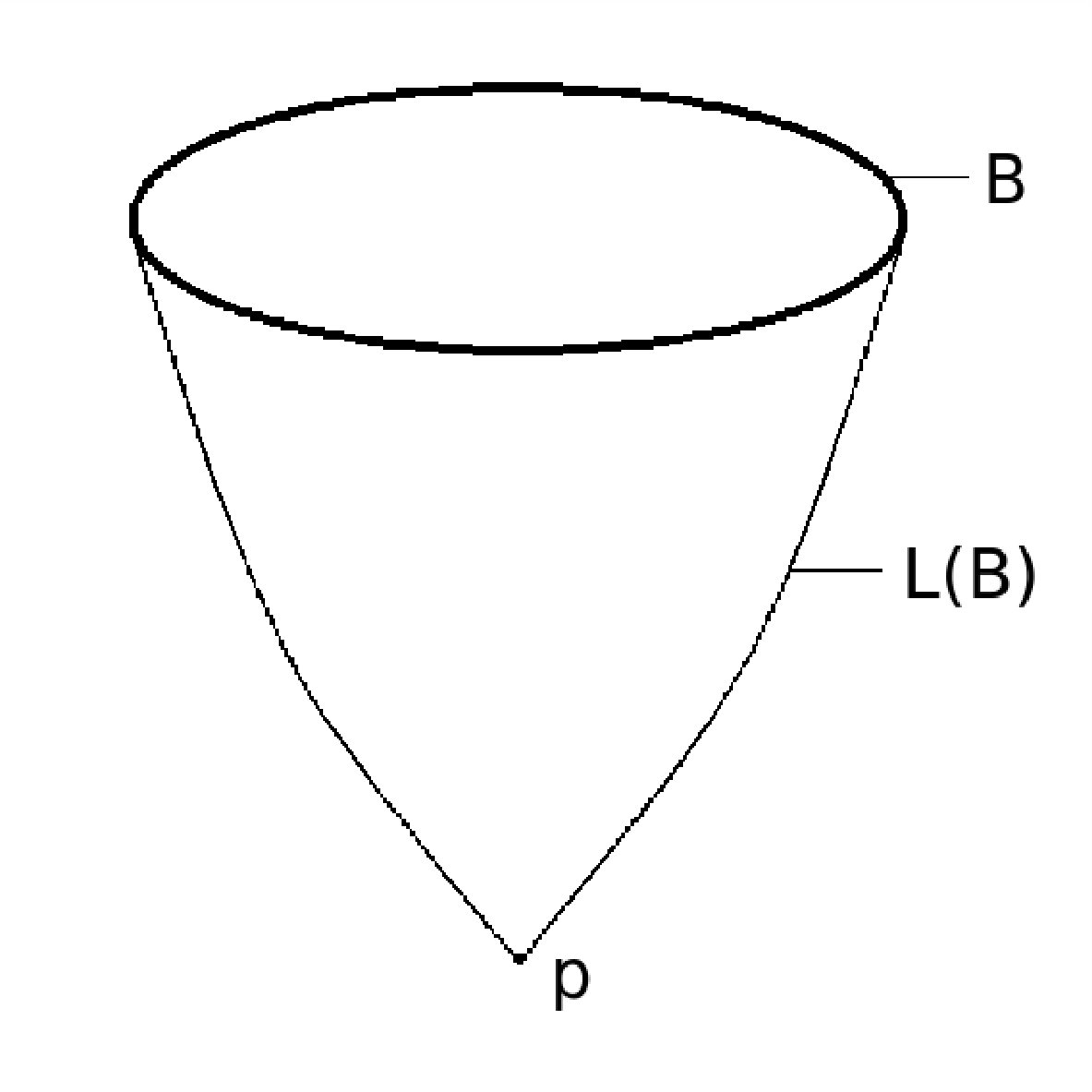}
\caption{L(B) is the portion of the future null cone of a point
$p$ up to a cross-section B such that the expansion of future directed
null rays is non-negative between $p$ and B. L(B) is a special
case of a `light sheet' that features in a more general statement of Bousso's
covariant entropy bound.}
\end{center}
\end{figure}

The simplest version of this conjecture can be stated as follows.
Let $(M,g_{ab})$ be a smooth classical space-time and let $p$ be a
point in it. Consider the future null cone of $p$ and let B be any
2-sphere cross section of the null cone such that the expansion of
null rays between $p$ and B is always non-negative. Denote this
portion of the null cone by L(B). Then the Bousso's conjecture
states that the entropy flux across L(B) is bounded by $A_{\rm
B}/4\lp^2$ where $A_{\rm B}$ is the area of B: $\int_{\rm L(B)} S^a
dA_a \le A_{\rm B}/4\lp^2$, where $S^a$ is the entropy-flux 4-vector
field. This statement has three curious features.  First, it
requires a smooth classical space-time $(M, g_{ab})$, without which
one cannot even define null rays and their expansion. Second, it
requires that the notion of the entropy-flux $S^a$ be well defined.
Finally, although it makes a crucial use of a classical metric
$g_{ab}$, since its statement features Planck length, the conjecture
is a statement about quantum gravity rather than classical general
relativity. Indeed, in the limit $\hbar \rightarrow 0$ the bound
becomes infinite and the conjecture trivializes. In spite of these
peculiarities, however, the conjecture holds in a large number of
circumstances \cite{bousso2}. Consequently, in some particle physics
circles it is taken to have deep and fundamental significance.
Indeed it is sometimes stated that the covariant entropy bound is to
quantum gravity what the equivalence principle was to general
relativity; it should be an essential building black of any
satisfactory quantum gravity theory.

Let us examine the bound in the simplest cosmological setting in
classical general relativity: the k=0, $\Lambda$=0 FRW model, filled
with radiation. Suppose the point $p$ is taken to lie on a
homogeneous slice $\tau=\tau_i$ and the surface B to lie on the
slice $\tau=\tau_f$. Then, a straightforward calculation enables one
\cite{awe1} to establish the following relation
\be \frac{S}{A_{\rm B}} = \frac{\lp^2}{6}\,\,
(\frac{2}{45\pi})^{1/4}\,\,
 \frac{\sqrt{\lp}}{\sqrt{\tau_f}}\,\, \left( 1 -
 \sqrt{\frac{\tau_i}{\tau_f}} \right) \ee
By plugging-in numbers, one finds that the bound $S/A_{\rm B} \le
0.25/\lp^2$ holds if $\tau_f > 0.1 \lp$. Since $\tau_i < \tau_f$,
the bound is in fact respected unless we choose the initial point
$p$ to lie \emph{extremely} close the big-bang singularity and
terminate the light cone \emph{very} close to $p$. However, the
bound \emph{is} violated in the immediate future of the singularity.
Usually it is argued that in this region quantum gravity effects
should dominate and hence the failure of the bound cannot be taken
seriously; quantum corrections will intervene and save the bound.

Is this really the case? A priori this possibility seems almost
impossible to test because of the three peculiar features I
mentioned above. On the one hand, quantum gravity effects are
essential. On the other, if they lead to significant fluctuations in
the space-time metric, one loses even the formulation of the
conjecture! Surprisingly, LQC provides an ideal arena to test this
possibility because of the following two features. First, quantum
corrections can indeed intervene; they are so strong that the
singularity is resolved. Second, in spite of thes strong effects,
there is a smooth metric even at the bounce at which the quantum
state is sharply peaked. This is not the metric given by Einstein's
equations but rather by the LQC effective equations which
incorporate the quantum corrections that unleash the tremendous
repulsive force of quantum geometric origin. Still, since there is a
smooth metric, the conjecture is meaningful and can therefore be
tested. A detailed calculation \cite{awe1} shows:
\be  \frac{S}{A_{\rm B}}\,\,  <  \,\, \f{0.244}{\lp^2}\, . \ee
Thus in LQC the bound is in fact respected. Although the calculation
has several limitations (see \cite{awe1}), it nonetheless provides
an interesting and completely unforeseen convergence of very
different ideas related to quantum gravity.

This result suggests the following overall viewpoint on the entropy
bound. Recall that the bound is strongly motivated by the
generalized second law of thermodynamics (which also does not have a
sharp, definitive formulation). Now, already the standard second law
of thermodynamics is a deep fact of Nature but it has a `fuzziness'
which is not shared by other deep laws such as energy-momentum
conservation. In particular, the second law requires a coarse
graining in an essential way. It is not a statement about the
evolution of micro-states; in a fundamental theory their dynamics is
always time reversible (leaving aside, for simplicity, quantum
measurements). Rather, it is a statement about how the number of
micro-states compatible with a pre-specified coarse graining changes
in time. For specific processes, the increase of entropy can be
calculated using statistical mechanics. But this entropy has little
relevance to the fundamental dynamics of micro-states and is not an
\emph{input} in the construction of statistical mechanics. In the
same vein, it seems unlikely that covariant entropy bounds would be
essential ingredients in the \emph{construction} of a quantum theory
of gravity. It seems more natural to expect the covariant entropy
bound should \emph{emerge} from a fundamental quantum gravity theory
under suitable conditions. Returning to the LQC calculation, the
distinguishing feature of LQC is the underlying quantum geometry.
The covariant entropy bound was never an input or even a motivation.
Rather, it emerged on the quantum corrected semi-classical
space-time that results from the LQC dynamics.

\section{Quantum Dynamics}
\label{a2}

This appendix is divided into four parts. In the first I discuss how
the Hamiltonian constraint is handled in LQC. In the second, I point
out that, although the difference between the `$\mu_o$ and $\bar\mu$
schemes' is quite subtle, the resulting quantum constraints have
very different predictions. Quantum dynamics resolves the
singularity in both cases but the $\mu_o$ scheme leads to
inadmissible infrared behavior. This difference has valuable lessons
for the construction of the quantum Hamiltonian constraint in full
LQG. In the third, I  summarize the analytical results that provide
a precise sense in which the singularity is resolved in the
$\bar\mu$ scheme. In the fourth, I will conclude by providing a
precise relation between the results of LQC and those of the \WDW
theory.

\subsection{How is dynamics handled in LQC?}
\label{a2.1}

The Hamiltonian constraint has the following form in full LQG:
\be C_{\rm H} \,\,\sim \,\, \underbrace{(\epsilon^{ij}{}_k E^a_i
E^b_j/\sqrt{q})}_{\rm Triads}\, \underbrace{F_{ab}^k}_{\rm
connection} \ee
where the first term is determined by spatial triads $E^a_i$ and the
second by the gravitational spin connection $A_a^i$. To define
quantum dynamics, we have to first find the corresponding operator
$\h{C}_{\rm H}$ on the kinematical Hilbert space $\Hk$ of LQC.
Thanks to a general procedure introduced by Thiemann, we know how to
handle the first term involving triads. That procedure has a direct
analog in LQC (see, e.g., \cite{abl}). However, the second term
poses a challenge: Although the curvature $F_{ab}{}^k$ has a simple
expression in terms of $A_a^i$, as we saw in section \ref{s4}, there
is no operator corresponding to the connection either in LQG or LQC.

But the holonomy operators are well defined in both theories. Indeed
they are the fundamental/elementary configuration operators on
$\Hk$. Now, to quantize a function $f$ on any phase space, the
standard strategy in quantum physics is to first express $f$  as a
suitable function of the elementary configuration and momentum
variables and then replace each of these variables by its quantum
analogs. Modulo factor ordering, this procedure yields the desired
quantum operator $\hat{f}$. It is natural to follow this procedure
to obtain the operator $\hat{F}_{ab}{}^k$. The first step is
straightforward. For, the $a$-$b$ component of curvature at a point
$p$ can be expressed in terms of the holonomy $h_{\rm \Box}$ around
a closed loop $\Box$ in the $a$-$b$ plane at $p$ as a limit,
\be \label{F} F_{ab}^k = -2 \lim_{{\rm Ar}_\Box \rightarrow 0}\,\,
\f{{\rm Tr}( h_{\Box_{ab}} -1 )\tau^k}{{\rm Ar_\Box}}\, , \ee
in which the area of the 2-surface enclosed by the loop $\Box$ tends
to zero. In the second step we can to replace the holonomy
$h_{\Box}$ by the operator $\h{h}_{\Box}$ on $\Hk$. However, as we
shrink the loop, the limit of this operator does not exist on $\Hk$.
This is not accidental but is tied to the fundamental feature of the
unique Hilbert space $\Hk$ we are led to by the requirement of
diffeomorphism covariance \cite{lost}. Indeed, had this limit
existed, we would have had a local operator corresponding to the
gravitational connection $A_a^i$. Now, in full LQG, this difficulty
can be bypassed by first solving the diffeomorphism constraint and
then working on the space $\H_{\rm diff}$ of diffeomorphism states
rather than $\Hk$. On these diffeomorphism invariant states $\Psi$,
we can easily shrink the loop because, once the loop is sufficiently
small, $\h{h}_{\Box}\, \Psi$ remains invariant under the operation
of shrinking the loop further.

In LQC, on the other hand, since we have eliminated the
diffeomorphism constraint by gauge fixing, we have to work directly
on $\Hk$. Therefore, to define $\h{F}_{ab}^k$, one adopts the
following working strategy \cite{abl}. Recall that in  LQG the
connection operator is not well-defined because the holonomy fails
to be weakly continuous in the loop and it is precisely this lack of
continuity that leads one to quantum geometry in which geometric
operators ---particularly areas of surfaces--- have purely discrete
eigenvalues. This interplay is taken to suggest that we have to take
the quantum nature of geometry seriously. We cannot shrink the loop
to zero area continuously. Rather, we should define $\h{F}_{ab}^k$
simply by evaluating $\h{h}_{\Box}$ in the operator analog of
(\ref{F}) around a loop $\Box$ which encloses an area equal to the
smallest non-zero eigenvalue $\Delta \lp^2$ of the area operator,
called the \emph{area gap}. The resulting operator is an acceptable
quantization of $\h{F}_{ab}{}^k$ because is has the correct
classical limit $F_{ab}^k$ in a precise sense. However, at a
fundamental level the operator has a \emph{Planck scale
non-locality}. It is only in the classical limit that it reduces to
the familiar local form $F_{ab}{}^k = 2 \partial_{[a}A_{b]} +
\epsilon_{ij}{}^k A_a^i A_b^j$.

The technical implementation of this idea involves two subtleties
that were overlooked in the beginning. In the first implementation
\cite{abl}, the loop $\Box$ was chosen so that the area it encloses
with respect to the \emph{fiducial} metric is $\Delta\,\lp^2$. This
is called \emph{the $\mu_o$ scheme}. It led to a quantum constraint
that resolves the singularity \cite{aps2}. However, as we will see
in appendix \ref{a2.2}, it has severe limitations. In particular, it
predicts deviations from general relativity even in certain `tame'
situations. However it was soon realized that these difficulties go
away if, in the construction of $\hat{F}_{ab}{}^k$ the area enclosed
by $\Box$ is computed using the \emph{physical} metric $q_{ab}$
rather than the fiducial metric $\q_{ab}$ \cite{aps3}. This switch
is conceptually well motivated since quantization of area
eigenvalues refers to the actual metric used in the LQG kinematics,
not an auxiliary or a fiducial one. This strategy is referred to in
the literature as \emph{the $\bar\mu$ scheme}. While the technical
implementation of this strategy is rather subtle, the final
Hamiltonian constraint (\ref{hc4}) is a rather simple difference
operator in the volume variable $v$. As one might expect, the step
size of the operator is dictated by the area gap.%
\footnote{In the $\mu_o$ scheme,  the `lattice' has uniform spacing
in $p$, the variable which measures the area of each surface of the
fiducial cell while in the $\bar\mu$ scheme it has uniform spacing
in volume $v$. Since $v \sim p^{3/2}$, the dynamics of $\bar\mu$
scheme cannot be supported by alternative LQC kinematics of the type
proposed by Velhinho \cite{vel}.}
In the limit in which the gap goes to zero, this quantum constraint
(\ref{hc4}) reduces to the Hamiltonian constraints (\ref{wdw0}) and
(\ref{wdw1}) of the \WDW theory in the sense spelled out in appendix
\ref{a2.4}.

The second subtlety is less significant. In much of the literature
\cite{aps3,apsv,warsaw1,acs,cs} the area gap $\Delta \lp^2$ was
taken to be $2\sqrt{3}\pi \gamma \lp^2$, the lowest of all non-zero
eigenvalues of the area operator on $\Hk$ (where $\gamma$ is the
Barbero-Immirzi parameter). However, it was later realized
\cite{awe2,awe1} that, if one uses a semi-heuristic correspondence
between LQG and LQC states, LQG eigenstates with these eigenvalues
will not appear in quantum geometries needed to represent
homogeneous classical metrics. On states that can feature, the
minimum non-zero eigenvalue is \emph{twice as large}. This shift has
no qualitative effect on the earlier results in the $\bar\mu$ scheme
(e.g., those in \cite{aps3,apsv,warsaw1,acs,cs}). The only
difference is a factor of 2 in some expressions. In particular, the
critical density $\rcr$ at which the quantum bounce occurs in these
models goes inversely as $\Delta$, whence in the earlier literature,
$\rcr$ was $\sim 0.82\rho_{\rm{Pl}}$ rather than $\sim 0.41\rho_{\rm
Pl}$.

Note that I have referred to this construction as a `working
strategy'. This is because one has to `parachute' into LQC certain
important facts about quantum geometry from the kinematics of full
LQG. This is analogous to Bohr's treatment of the hydrogen atom
where, in retrospect, one can say that quantization of angular
momentum is `parachuted' into the model from full quantum mechanics.
With incorporation of this one essential feature, the simple Bohr
model is capable of giving energy levels to a remarkable accuracy.
However, at a fundamental level, one has to use full quantum
treatment a la Pauli or Schr\"odinger. Similarly, in quantum
cosmology a more fundamental construction will emerge only from a
better understanding of the detailed relation between LQG and LQC.

\subsection{What are the differences between the $\mu_o$ and $\bar\mu$
 quantum theories?}
\label{a2.2}

Recall, first, that in the k=0 model on $\mathbb{R}^3$, it is often
convenient to introduce a fiducial metric $q^o_{ab}$ and label the
physical metric $q_{ab}$ via a scale factor $a(t)$: $q_{ab} =
a^2(t)\, q^o_{ab}$. If we rescale the fiducial metric via $q_{ab}^o
\rightarrow \alpha^2\, q^o_{ab}$, the descriptor $a$ of the physical
metric $q_{ab}$ changes via $a \rightarrow \alpha^{-1}a$ even though
$q_{ab}$ itself is untouched. Of course physics can not change under
such rescalings. The construction of a Hamiltonian or Lagrangian
framework requires an additional auxiliary structure because one has
to perform integrals of physical fields over space and these diverge
because of homogeneity. Therefore, one has to introduce a fiducial
cell $\mathcal{C}$ and restrict all integrals to $\mathcal{C}$. In
classical general relativity, one can work directly with field
equations and the auxiliary $\mathcal{C}$ is unnecessary. However it
is essential for passage to quantum theory, irrespective of whether
one uses path integrals or canonical quantization.

Thus, in any quantum cosmology on non-compact, spatially homogeneous
manifolds, there are two independent auxiliary structures, the
metric $q^o_{ab}$ and the cell $\mathcal{C}$. Each can be changed
keeping the other fixed. At the classical level, the final physical
results do not depend on the specific choices one makes to construct
the Hamiltonian theory. But a priori there is no guarantee that
results of the quantum theory will continue to enjoy this feature.
If they do not in a specific quantum theory, that theory cannot be
physically viable. The first contrast between the two schemes is
that while in the $\mu_o$ scheme the final results of the quantum
theory depend on
these auxiliary structures%
\footnote{ It is often possible to hide the dependence on the choice
of $q^o_{ab}$ by shifting it to the choice of the cell $\mathcal{C}$
or vice versa but of course physics has to be invariant with respect
to both these choices.},
in the $\bar\mu$ scheme they do not.

Since a cell $\mathcal{C}$ is in any case necessary to the
Hamiltonian framework, one can make kinematics manifestly
independent of the fiducial metric $q^o_{ab}$ by rescaling the
canonical variables by suitable powers of the volume $V_o$ of
$\mathcal{C}$ with respect to $q^o_{ab}$ \cite{abl}. Then the
canonical commutation relations, and expressions of the elementary
operators are all independent of the choice of $q^o_{ab}$. Thus, for
each choice of $\mathcal{C}$ one constructs a  quantum theory. Now,
suppose we are given a classical solution $(a(t), \phi(t))$. Within
each quantum theory, one can construct semi-classical states peaked
at this solution at a late time when curvature is low and evolve it
back in time. One can then ask: when do quantum effects become
important? Unfortunately, in the $\mu_o$ scheme, the answer depends
on the choice of the cell $\mathcal{C}$! (See appendix B.2 in
\cite{aps3}). In particular, the matter density $\rho_b$ at the
bounce point is given by $\rho_b = (3/8\pi G
\gamma^2\mu_o^2)^{3/2}\,\, (\sqrt{2}/p_{(\phi)}) \equiv C
/p_{(\phi)}$ where by inspection the constant $C$ does not depend on
the choice of $\mathcal{C}$ (or $q^o_{ab}$). However, since $p_\phi
= \dot{\phi}\,V$ where $V$ is the physical volume of $\mathcal{C}$,
$\rho_b$ decreases linearly with the size of $\mathcal{C}$. Thus, in
the $\mu_o$ scheme, even the density at the bounce point fails to
have a physical meaning. By contrast, in the $\bar\mu$ scheme,
$\rho_b$ is universal, independent of the choice of $\mathcal{C}$.

A second difference is that the $\mu_o$ scheme has serious problems
with recovering general relativity in the low curvature limit while
the $\bar\mu$ scheme is free of this drawback. To see this, let us
fix a cell $\mathcal{C}$ and examine the resulting quantum theory.
As we just saw, in the k=0, $\Lambda$=0 case, the matter density at
the bounce point is given by  $\rho_b = C/p_{(\phi)}$. Now, $p_\phi$
is a constant of motion and larger its value, more semi-classical
the state is. (For example, as we saw in section \ref{s5}, in the
k=1 case the maximum radius of the universe goes as
$p_{(\phi)}^{3/2}$.) So, $\rho_b$ can be made as low as we want
simply by increasing $p_{(\phi)}$. In particular, there are states
which are extremely semi-classical at late times which, when evolved
back undergo a quantum bounce at density of water! A gross violation
of general relativity at such densities is clearly unphysical. In
the $\bar\mu$ scheme on the other hand, $\rho_b \approx
0.41\rho_{\rm Pl}$ is universal, irrespective of how large
$p_{(\phi)}$ is.

A third difference between the two schemes arises in presence of a
cosmological constant. If $\Lambda \not=0$, problems with the
infra-red behavior also arise in the $\mu_o$ scheme \emph{well away
from the bounce}. Even at late times, the quantum theory exhibits
large departures from general relativity in low curvature regimes.
Again, this cannot happen in the $\bar\mu$ scheme. It is quite
remarkable that a subtle but physically well-motivated correction in
the construction of the quantum field strength operator cures all
the conceptual and physical pathologies of the $\mu_o$ scheme
without affecting the singularity resolution. It is a marvellous
example of the deep harmony between physics and mathematics.

The difference between the two schemes also has important lessons
for full LQG. Recently, Giesel et al \cite{tt} have pioneered a
program to deparametrize general relativity along the lines of Brown
and Kucha\v{r} \cite{bk}, making the treatment of constraints in LQG
conceptually similar to that in LQC. This is a notable advance.
However, quantization of the Hamiltonian constraint continues to
face a host of ambiguities in full LQG. The fact that the apparently
natural $\mu_o$ scheme is not viable even in the simplest LQC models
suggests that a much greater control on the semi-classical
predictions is needed to weed out ambiguities. Specifically, an
understanding the relation between choices made in the quantization
of the Hamiltonian constraint and the behavior of the resulting
physical states in the semi-classical regime would go a long way to
reducing the large apparent freedom one is currently faced with.

Finally, in this appendix I have spelled out differences between the
$\mu_o$ and $\bar\mu$ schemes in general terms because the more
complete discussion of \cite{aps3,acs,cs} is rather technical. As a
result, it appears not to have been fully absorbed by the community
and some authors continue to treat the two schemes on an almost
equal footing.

\subsection{Are there analytical results that elucidate
properties of the quantum bounce?}
 \label{a2.3}

There are two analytical results that are particularly interesting
because they zero-in on the key features of the bounce: the first
provides an upper bound on the spectrum of the matter density
operator in LQC and the second shows that the `bounce' is not
restricted to states which are semi-classical at late times but
occurs for all states in the domain of the volume operator.

In the first papers \cite{aps1,aps2} on the bounce, the analysis
began with the choice $N=1$ of the lapse function. In the classical
theory the corresponding Hamiltonian constraint generates evolution
in proper time. One then found the quantum operator corresponding to
this constraint. But as we saw, physical interpretation of the
theory is simplest if one uses the scalar field $\phi$ as internal
time. The corresponding lapse is $N = V$, the volume of the fiducial
cell with respect to the physical
metric $q_{ab}$.%
\footnote{This corresponds to working in the harmonic-time gauge: on
space-time $ds^2 = - a^3(T)V_o dT^2 + a^2(T) d\vec{x}^2$, the time
coordinate $T$ satisfies the wave equation and the scalar field
$\phi(T)$ is given by $\phi(T) = (p_{(\phi)}/V_o) T$ (where $V_o$ is
the volume of the cell $\mathcal{C}$ with respect to the fiducial
metric $q^o_{ab}$ and $p_\phi$ is a constant of motion).}
One can begin with this lapse already in the classical theory and
then quantize. If one does so, the constraint is simpler because it
does not involve any inverse powers of $V$ (even if one were to
include dust, radiation or other kinds of matter used in the
cosmology literature). This strategy was not adopted in the first
papers because, if the underlying 3-manifold is non-compact, the
strategy does not naturally generalize to inhomogeneous contexts.
However, it has some conceptual and technical advantages in
homogeneous cosmologies. In fact the quantum constraint coincides
with that of the so-called simplified LQC --called sLQC in the
literature-- \emph{which can be solved exactly analytically}
\cite{acs}.\, sLQC itself was first arrived at by first using $N=1$,
going to the quantum theory and then making certain simplifications
in the resulting quantum constraint \cite{acs}. This round-about
procedure is not necessary. One arrives at the same exactly soluble
quantum theory if one formulates the classical theory with $N=V$,
i.e., with the scalar field as internal time, and then proceeds with
quantization.

Since this theory is exactly soluble, one can establish several
interesting results analytically. First consider the matter density
operator $\hat \rho|_\phi = \textstyle{\frac{1}{2}}\,
(\hat{V}_\phi)^{-1}|_\phi\,\, \hat{p}_{(\phi)}^2\,\,
(\hat{V}_\phi)^{-1}|_\phi$ at the instant $\phi$ of the internal
time. One can show that its spectrum has an absolute upper bound
$\rho_{\rm sup}$ \emph{on the physical Hilbert space} \cite{acs}:
\be \rho_{\rm sup} = \f{\sqrt{3}}{16\pi^2\gamma^3G^2\hbar}
\,\,\approx\, 0.41 \rho_{\rm Pl}.\ee
Since the inverse volume operator $\widehat{V^{-1}}$ is bounded
above, already on the kinematic Hilbert space $\Hk$, $\hat\rho$ is
bounded above \emph{provided one restricts oneself to the subspace
of the Hilbert space spanned by the eigenstates of
$\hat{p}_{(\phi)}$ with eigenvalues below any fixed $p_{(\phi)}^o$}.
This kinematical result is sometimes used to argue that singularity
is resolved (see, e.g., \cite{mv}). However, since the full spectrum
of $\hat{p}_{(\phi)}$ is \emph{un}bounded above, so is $\hat\rho$ on
$\Hk$. The fact that it admits an absolute upper bound on $\Hp$ does
not follow from the boundedness of $\widehat{V^{-1}}$; it is a
highly non-trivial consequence of \emph{quantum dynamics.} It
provides a clear-cut sense in which the singularity is resolved in
LQC.

It is natural to ask: How good is this bound $\rho_{\rm sup}$? Is it
actually achieved? Using states which are Gaussians at late times,
one can show that the expectation values $\rho_{\rm bounce}$ of the
density operator at the bounce point approach arbitrarily close to
$\rho_{\rm sup}$:
\be\rho_{\rm bounce} = \rho_{\rm sup}\, \left[ 1 - O
\big(\frac{G\hbar^2}{p_{(\phi)}^2 + (\Delta
p_{(\phi)})^2}\big)\right]\ee
where $\Delta p_{(\phi)}$ is the dispersion in $\hat{p}_{(\phi)}$
Thus, $\rho_{\rm sup}$ is indeed the lowest upper bound on the
spectrum of the density operator. Chronologically, $\rho_{\rm
bounce}$ was first found in numerical simulations. It turned out to
be remarkably robust in the sense that it is insensitive to modeling
of the semi-classical states in three different ways and does not
alter when a non-zero cosmological constant is included
\cite{aps3,bp}or when the analysis is extended to the k=1 case
\cite{apsv}. This robustness suggested that there must be an
analytical result that $\hat\rho$ is a bounded operator, which
indeed turned out to be the case. This is a nice example of the
synergistic interplay between numerical and analytical work.

Finally, we can ask if the quantum bounce is restricted only to
states which are semi-classical at late times. On general states,
one has to first define what one means by a bounce. A natural
strategy is to examine expectation values of the volume operator. In
the exactly soluble model, one finds \cite{acs} that for \emph{any
state in the domain of the volume operator} we have:
\be (\Psi,\, \hat{V}_\phi \Psi)_{\rm Phy} = V_+ e^{\sqrt{12\pi
G}\phi} + V_-  e^{-\sqrt{12\pi G}\phi}\ee
where $V_\pm$ are determined by the `initial data' $\Psi(v, \phi_o)$
for the state at any $\phi_o$. Therefore, it follows that the
expectation value is well-defined for all finite times $\phi$ but
grows unboundedly as $\phi \rightarrow \pm \infty$. Hence it must
attain a minimum. Indeed, the minimum is given by:
\be V_{\rm min} = \sqrt{(V_- V_+)}\ee
and occurs at time $\phi_{\rm bounce} = \frac{1}{2\sqrt{12\pi
G}}\, (\ln V_- - \ln V_+)$.

 \subsection{What is the precise relation between the LQC and the
\WDW dynamics?} \label{a2.4}

This question has been analyzed in detail for the exactly soluble
theory (i.e., the k=0, $\Lambda$=0 model) I just discussed in
appendix \ref{a2.3}. Let us start with a physical state at a late
time $\phi=\phi_o$ when the matter density and curvature are low,
evolve it using LQC and the \WDW theory and compare the results. In
the LQC evolution, the area gap $\Delta \lp^2$ plays a key role
while in the \WDW theory, it is effectively zero. To compare the two
evolutions, let us not fix $\Delta\lp^2$ to its numerical value
$4\sqrt{3}\pi \gamma\lp^2$ used in LQC but allow $\Delta$ to vary so
we can take the limit $\Delta \rightarrow 0$. Then:

\begin{itemize}
\item Certain predictions of sLQC approach those of the
WDW theory as $\Delta$ goes to zero. \\
That is, given a semi-infinite `time' interval $I_\phi$ and an
$\epsilon >0$, there exists a $\delta>0$ such that $\forall
\Delta <\delta$, `physical predictions of the two theories are
within $\epsilon$ of each other' if restrict $\phi$ to lie in
$I_\phi$.

\item However, approximation is \emph{not} uniform. The WDW theory
is \emph{not} the limit of LQC as $\Delta$ goes to zero. \\
Thus, given $N>0$ however large, there exists a $\phi$ such that
$\langle \hat{V}_{\phi} \rangle_{\rm sLQC} -\langle \hat{V}_{\phi}
\rangle_{\rm WDW} > N$. In this sense, LQC is \emph{fundamentally}
discrete.

\end{itemize}
Thus the relation is somewhat more subtle than one might have first
thought. If we are interested in physical predictions either to the
past or to the future of any fixed time $\phi =\phi_o$, LQC does
reduce to the \WDW theory in the limit in which the area gap is
taken to zero. But the \WDW theory is \emph{not} the `continuum
limit' of LQC because if we include the full time interval, there
are many physical predictions of LQC which will not reduce to those
of the \WDW theory. In fact, as a full fledged theory, LQC does not
admit a continuum limit; it is a fundamentally discrete theory.

For more precise and detailed statements, see \cite{acs}.

\end{appendix}


%

\end{document}